\documentclass[fleqn,usenatbib]{mnras}

\usepackage{newtxtext,newtxmath}

\usepackage[T1]{fontenc}

\DeclareRobustCommand{\VAN}[3]{#2}
\let\VANthebibliography\thebibliography
\def\thebibliography{\DeclareRobustCommand{\VAN}[3]{##3}\VANthebibliography}

\usepackage{graphicx}
\usepackage{amsmath}
\usepackage{xspace}
\usepackage{siunitx}
\usepackage{booktabs}
\usepackage{bold-extra}
\usepackage{orcidlink}
\usepackage[capitalise]{cleveref}

\crefname{equation}{equation}{equations}
\Crefname{equation}{Equation}{Equations}

\newcommand*{\ywinprep}{Wang, Y.\ et al.\ (in prep.)\xspace}
\newcommand*{\ywinprepp}{Wang, Y.\ et al.\ in prep.\xspace}

\newcommand*{\Bhatulap}{Bhatula et al.\ in prep.\xspace}

\newcommand*{\mr}[1]{{\mathrm{#1}}}
\newcommand*{\DM}{\ensuremath{\mr{DM}}\xspace}
\newcommand*{\DMh}{\ensuremath{\DM_\mr{host}}\xspace}
\newcommand*{\DMd}{\ensuremath{\DM_\mr{d}}\xspace}
\newcommand*{\DMMW}{\ensuremath{\DM_\mr{MW}}\xspace}
\newcommand*{\DMhalo}{\ensuremath{\DM_\mr{MWhalo}}\xspace}
\newcommand*{\DMX}{\ensuremath{\DM_\mr{X}}\xspace}
\newcommand*{\DMs}{\ensuremath{\DM_\mr{src}}\xspace}

\newcommand*{\Dhs}{\ensuremath{D_\mr{host,src}}\xspace}
\newcommand*{\Ds}{\ensuremath{D_\mr{src}}\xspace}
\newcommand*{\Dh}{\ensuremath{D_\mr{host}}\xspace}
\newcommand*{\thh}{\ensuremath{\theta_\mr{host}}\xspace}

\newcommand*{\nusc}[1][]{\ensuremath{\nu_{\mr{s}{#1}}}\xspace}
\newcommand*{\taus}[1][]{\ensuremath{\tau_{\mr{s}{#1}}}\xspace}
\newcommand*{\SNR}{\ensuremath{S/N}\xspace}

\newcommand*{\tool}[1]{{\textsc{#1}}}
\newcommand*{\FRBo}{FRB\,240210D\xspace}
\newcommand*{\FRBt}{FRB\,240312D\xspace}
\newcommand*{\FRBfaber}{FRB\,221219A\xspace}
\newcommand*{\FRBshin}{FRB\,200723B\xspace}
\newcommand*{\FRBYrep}{FRB\,240216A\xspace}

\newcommand*{\HII}{H\,\textsc{ii}\xspace}
\newcommand*{\OIIIa}{[O\,\textsc{iii}]\,$\lambda 4959$\xspace}
\newcommand*{\OIIIb}{[O\,\textsc{iii}]\,$\lambda 5007$\xspace}
\newcommand*{\NIIa}{[N\,\textsc{ii}]\,$\lambda 6548$\xspace}
\newcommand*{\NIIb}{[N\,\textsc{ii}]\,$\lambda 6584$\xspace}
\newcommand*{\SIIa}{[S\,\textsc{ii}]\,$\lambda 6717$\xspace}
\newcommand*{\SIIb}{[S\,\textsc{ii}]\,$\lambda 6731$\xspace}
\newcommand*{\OI}{[O\,\textsc{i}]\,$\lambda 6300$\xspace}
\newcommand*{\HeI}{He\,\textsc{i}\,$\lambda 5876$\xspace}
\newcommand*{\ArIII}{[Ar\,\textsc{iii}]\,$\lambda 7135$\xspace}

\DeclareSIUnit \pc {pc}
\DeclareSIUnit \Mpc {\mega\pc}
\DeclareSIUnit \kpc {\kilo\pc}
\DeclareSIUnit \erg {erg}
\DeclareSIUnit \Jy {Jy}
\DeclareSIUnit \dmu {\pc\per\cm\cubed}
\DeclareSIUnit \yr {yr}
\DeclareSIUnit \arcsecletter {arcsec}
\DeclareSIUnit \Msol {\ensuremath{\mr{M}_\odot}}
\DeclareSIUnit \gauss{G}
\DeclareSIUnit \sky {sky}
\DeclareSIUnit \FRBs {FRBs}
\DeclareSIUnit \angstrom {\text {Å}}

\makeatletter
\newcounter{affilCount}
\newcommand{\affn}[1]{%
  \@ifundefined{afn@#1}%
    {\stepcounter{affilCount}%
     \expandafter\xdef\csname afn@#1\endcsname{\number\value{affilCount}}}%
    {}%
  \csname afn@#1\endcsname%
}
\makeatother

\graphicspath{{./}{figures/}}

\title[Highly Scattered FRBs]{Two new highly scattered fast radio bursts: evidence for scatter broadening by the circumsource medium}

\author[J. N. Jahns-Schindler et al.]{Joscha N.\ Jahns-Schindler \orcidlink{0000-0003-4193-6158},$^{\affn{Swin}}$\thanks{E-mail: jjahnsschindler@swin.edu.au}
Keith W. Bannister \orcidlink{0000-0003-2149-0363},$^{\affn{CSIRO},\affn{Syd}}$
Adam~T.~Deller \orcidlink{0000-0001-9434-3837},$^{\affn{Swin},\affn{OzG}}$
Xinping Deng \orcidlink{0009-0002-1153-0077},$^{\affn{CB}}$\newauthor
Marcin Glowacki \orcidlink{0000-0002-5067-8894},$^{\affn{Edin},\affn{Cape}}$
Alexa~C.~Gordon \orcidlink{0000-0002-5025-4645},$^{\affn{NU}}$
Vivek Gupta \orcidlink{0000-0001-9817-4938},$^{\affn{CSIRO}}$
Akhil Jaini \orcidlink{0000-0002-8987-1544},$^{\affn{Swin}}$
Clancy~W.~James \orcidlink{0000-0002-6437-6176},$^{\affn{ICRAR}}$\newauthor
Ilya S. Khrykin \orcidlink{0000-0003-0574-7421},$^{\affn{Val}}$
Yu Wing Joshua Lee \orcidlink{0009-0001-1631-7462},$^{\affn{Syd},\affn{OzG},\affn{ATNF}}$
J. Xavier Prochaska \orcidlink{},$^{\affn{UCSC},\affn{IPMU},\affn{NAOJ}}$
Hao Qiu \orcidlink{0000-0002-9586-7904},$^{\affn{SKAO}}$\newauthor
Hugh Roxburgh \orcidlink{0009-0001-6992-0898},$^{\affn{ICRAR}}$
Stuart D. Ryder \orcidlink{0000-0003-4501-8100},$^{\affn{macq},\affn{macqb}}$
Ryan~M.~Shannon \orcidlink{0000-0002-7285-6348},$^{\affn{Swin}}$
Tim Sprenger \orcidlink{0000-0002-1160-7276},$^{\affn{MPIfR}}$
Nicolas Tejos \orcidlink{0000-0002-1883-4252},$^{\affn{Val}}$\newauthor
Yuanming Wang \orcidlink{0000-0003-0203-1196},$^{\affn{Swin},\affn{OzG}}$ and
Ziteng Wang \orcidlink{0000-0002-2066-9823}$^{\affn{ICRAR}}$
\\
$^{\affn{Swin}}$Centre for Astrophysics and Supercomputing, Swinburne University of Technology, Hawthorn, VIC, 3122, Australia\\
$^{\affn{CSIRO}}$Australia Telescope National Facility, CSIRO, Space and Astronomy, PO Box 76, Epping, NSW 1710, Australia\\
$^{\affn{Syd}}$Sydney Institute for Astronomy, School of Physics, The University of Sydney, Sydney 2006, NSW, Australia\\
$^{\affn{OzG}}$OzGrav: The ARC Centre of Excellence for Gravitational Wave Discovery, Swinburne University of Technology, Hawthorn, VIC 3122, Australia\\
$^{\affn{CB}}$China-Brazil Belt and Road Joint Laboratory on Radio Astronomy Technology\\
$^{\affn{Edin}}$Institute for Astronomy, University of Edinburgh, Royal Observatory, Edinburgh, EH9 3HJ, United Kingdom\\
$^{\affn{Cape}}$Inter-University Institute for Data Intensive Astronomy, Department of Astronomy, University of Cape Town, Cape Town, South Africa\\
$^{\affn{NU}}$Center for Interdisciplinary Exploration and Research in Astrophysics (CIERA) and Department of Physics and Astronomy, Northwestern University,\\ \; Evanston, IL 60208, USA\\
$^{\affn{ICRAR}}$International Centre for Radio Astronomy Research, Curtin University, Bentley, WA, 6102, Australia\\
$^{\affn{ATNF}}$Australia Telescope National Facility, CSIRO, Space and Astronomy, PO Box 76, Epping 1710, NSW, Australia\\
$^{\affn{UCSC}}$Department of Astronomy and Astrophysics, University of California, Santa Cruz, CA 95064, USA\\
$^{\affn{IPMU}}$Kavli Institute for the Physics and Mathematics of the Universe (Kavli IPMU), 5-1-5 Kashiwanoha, Kashiwa, 277-8583, Japan\\
$^{\affn{NAOJ}}$Division of Science, National Astronomical Observatory of Japan,2-21-1 Osawa, Mitaka, Tokyo 181-8588, Japan\\
$^{\affn{SKAO}}$SKA Observatory, 26 Dick Perry Avenue, Kensington, WA 6151, Australia\\
$^{\affn{macq}}$School of Mathematical and Physical Sciences, Macquarie University, NSW 2109, Australia\\
$^{\affn{macqb}}$Astrophysics and Space Technologies Research Centre, Macquarie University, Sydney, NSW 2109, Australia\\
$^{\affn{MPIfR}}$Max-Planck-Institut für Radioastronomie, Auf dem Hügel 69, D-53121 Bonn, Germany\\
$^{\affn{Val}}$Instituto de F\'isica, Pontificia Universidad Cat\'olica de Valpara\'iso, Casilla 4059, Valpara\'iso, Chile
}

\date{Accepted XXX. Received YYY; in original form ZZZ}

\pubyear{\the\year{}}

\begin{document}
\label{firstpage}
\pagerange{\pageref{firstpage}--\pageref{lastpage}}
\maketitle

\begin{abstract}
We found two highly scattered Fast Radio Bursts (FRBs) during commissioning of the Commensal Realtime ASKAP Fast Transient COherent (CRACO) backend.
FRB\,240210D and FRB\,240312D have scattering times of $34\pm6$ and $300\pm48$\,ms, respectively, when scaled to 1\,GHz.
FRB\,240312D originates near a spiral arm of a face-on galaxy at a redshift of only 0.05.
Scintillation from a Milky Way screen constrains the distance of the scattering screen to $\sim 10$\,pc from the source.
FRB\,240312D is therefore the first highly scattered FRB where scattering screens in the host galaxy centre, a background galaxy, or intervening structures can all be excluded, leaving only the circumsource medium.
Integral field spectroscopy of the host reveals a Milky Way-like galaxy with a star-formation region at the FRB position.
We find refractive scattering in a pulsar wind nebula as the most likely scattering origin.
However, the explanation is not completely satisfactory as it requires a fine-tuned orientation.
Hence, additional theoretical studies under different FRB progenitor models are needed.
From the two FRBs, we calculate a total rate of $R_\mathrm{tot}=210^{+460}_{-180}\,\mathrm{events}\,\mathrm{sky}^{-1}\mathrm{day}^{-1}$
with durations between 55.2\,ms and 1\,s and above a fluence of 9\,Jy\,ms consistent with the rate of shorter FRBs.
This elevated rate suggests that the strong scattering seen in other FRBs likewise does not arise from chance-aligned sightlines, but is instead causally linked to the FRB sources.
\end{abstract}




\section{Introduction}
\label{sec:intro}

Fast Radio Bursts (FRBs) are extragalactic transients typically of \SI{\sim1}{\ms} duration \citep[see e.g.][for a review]{Petroff2019} observed at \SIrange{0.11}{8.4}{\GHz} frequencies \citep{Pleunis2021a,Snelders2023}.
At these frequencies, several propagation effects alter their time-frequency structure \citep[see e.g.][]{Lorimer2004}.
The three relevant to this work are dispersion, scatter broadening and scintillation.
Dispersion is the frequency-dependent delay of radio waves due to interaction with all free electrons along the path of propagation; it is quantified in terms of the Dispersion Measure (DM).
Scatter broadening and scintillation both stem from multipath propagation, or `scattering', caused by inhomogeneities in ionized intervening matter \citep{Scheuer1968,Rickett1969}. 
Scatter broadening refers to the temporal smearing of a burst due to refracted or diffracted waves travelling along different paths of various lengths.
Scintillation refers to patchy intensity observed in the spectrum, caused by interference of waves from different paths.
Scatter broadening and scintillation can originate from the same astrophysical `screen'.

In the simplest case of a single thin screen, the average delay or scattering time $\taus$ is related to the average bandwidth of scintles (intensity patches) or `scintillation bandwidth', $\nusc$, via $\nusc={1}/{2\pi\taus}$ \citep[see e.g.][]{Rickett1990,Pradeep2025}.
This relation can be used to identify several screens when scatter broadening and scintillation are observed in a source and do not follow the relation; most commonly these screens can be in the host galaxy and/or the Milky Way \citep{Masui2015, Sammons2023,Nimmo2025}.

Scattering is a very powerful probe of the enigmatic FRB origins.
Identifying screens in the host galaxy is the most sensitive way of constraining the emission region size \citep{Kumar2024,Nimmo2025}.
Moreover, scattering in the immediate source environment could give important clues about its properties.
Evidence for the presence of extreme magneto-ionic environments in the vicinity of FRBs comes from changing rotation measures \citep{Michilli2018,AnnaThomas2023} and compact persistent radio sources at the positions of FRBs \citep{Marcote2017,Niu2022,Bruni2025}.
Studies of scatter broadening showed variations in \taus of one repeater pointed to a dynamic, inhomogeneous plasma in the circumsource medium \citep{Ocker2023,Sand2023}.

Furthermore, scattering has been proposed as a tool for several astrophysical studies.
While the intergalactic medium (IGM) is not predicted to produce observable scattering \citep{Macquart2013,Luan2014,Xu2016} in line with measured upper limits \citep{Gupta2022a}, scattering can probe the presence of inhomogeneities in the circumgalactic medium of intersected galaxy haloes as well as the Milky Way halo \citep{Prochaska2019a, Ocker2021, Jow2024, MasRibas2025, Ocker2025}.
In the use of FRBs as cosmological probes through their DM \citep[see e.g.][]{Macquart2020, Connor2025}, the DM contribution from the host is one of the leading uncertainties.
Scattering has been proposed as an estimator for the host DM, which would increase the overall precision of FRBs \citep[see e.g.][]{Cordes2022,Ocker2022a}, although problems of detecting such a relation have been raised \citep{MasRibas2026}.

Because the signal-to-noise ratio (\SNR) decreases with the burst duration $w_t$ for a given fluence as $\SNR\propto w_t^{-1/2}$, it becomes increasingly difficult to detect longer bursts.
Since scattering broadens an FRB, it also reduces the detection \SNR.
This reduces the observed rate and could cause biases, e.g., in the observed host galaxy population and the observed locations in their hosts \citep[see e.g.][]{Gordon2025}.

As a result of the decreasing sensitivity with $w_t$, there are currently no observational constraints on the maximum width of FRBs \citep{James2026}.
A few not-so-fast radio bursts \citep[nsFRBs; $w_t>\SI{100}{\ms}$][]{Petroff2022} and highly scattered FRBs have been reported in the literature.
The class of `nsFRBs' is not strictly defined yet, and we propose to only use it for intrinsically long bursts with a fitted width $w_t>\SI{100}{\ms}$, where we define $w_t$ as 1.2$\times$ the full width at half maximum of the burst in the time series (this definition of burst width corresponds to the optimal boxcar search width for a Gaussian burst).
As a consequence, we will refer to bursts that have been scattered to long durations but are intrinsically smaller as `highly scattered' FRBs.
Following this definition, the first and only nsFRBs are the three nsFRBs found with Murriyang/Parkes by \citet{Crawford2022}.
The CHIME/FRB Catalog 2 \citep{CHIME2026} contains \num{\sim100} FRBs with fitted scattering times $>\SI{100}{\ms}$ at \SI{400}{\MHz}, 25 of which have been found at search boxcar widths $>\SI{100}{\ms}$.
The highest fitted width is \SI{27\pm3}{\ms}.
This illustrates the need for a definition like the above.
\citet{Faber2024} detected a highly scattered FRB and identified two intervening galaxy haloes as the most likely scatterers.
\citet{Shin2025} analysed the foreground of the most highly scattered CHIME FRB, with $\taus=\SI{1.025 \pm 0.009}{\s}$ at \SI{400}{\MHz}, and identified a cosmic filamentary structure; they nevertheless concluded that the scattering likely originates from the host galaxy.

Recently, the first searches for nsFRBs in the imaging domain were reported.
These have an advantage over searches in beamformed data because they are not subject to baseline variations and allow the use of the spacial information for radio frequency interference (RFI) rejection.
\citet{Wang2025} presented two detections with the CRACO upgrade that were found in a 110-ms pilot survey, with widths $<110.6$ and \SI{18.9}{\ms}.
They deduce a rate $\gtrsim1.7^{+6.5}_{-1.6}\times10^2$\,\si{\FRBs\per\sky\per\day} above a fluence of \SI{20.5}{\Jy\ms}.
In addition, \citet{Sherman2025} recently published upper limits of $<\SI{2}{\deg^{-2}\hour^{-1}}$ on Galactic sources from the Deep Synoptic Array-110.

In this work, we report the detections of two highly scattered FRBs found during commissioning of CRACO at \SI{13.8}{\ms} time resolution, which were subsequently localised to a few arcseconds.
The full survey and its other results will be described in a forthcoming paper (\ywinprepp).
We present the host galaxies, locations within their host galaxies, and constraints on a persistent radio counterpart of the closer FRB.
Moreover, we show that the scattering of \FRBt\footnote{Throughout the paper, we use a convention that omits the first two digits of the FRB name given by the Transient Name Server unless they are different from `20', i.e., we write \FRBt instead of FRB\,20240312D. This is supposed to aid visual recognition and memorisation.} most likely originates in the circumburst medium.
We discuss the implications for FRB origins, for the interpretation of other FRBs, and for scattering as a tool.

The paper is organized as follows.
Section~\ref{sec:method} describes the radio data, the scattering parameter extraction, and a rate estimate.
In Section~\ref{sec:results}, we present the results including the inferred rate.
Section~\ref{sec:followup} is about the multi-wavelength follow-up campaign, the host galaxies and our integral field spectroscopy of the \FRBt host.
Section~\ref{sec:discussion} holds our discussion of possible origins of the scattering, physical models that could provide the conditions for such screens, and the consequences of our findings.
We conclude in Section~\ref{sec:conclusion}.
Throughout the paper, we assume a $\Lambda$CDM cosmology with parameters from \citet{Planck2020}.

\section{FRB Observations and Analysis}
\label{sec:method}

\begin{figure*}
    \centering
    \includegraphics[width=0.49\linewidth]{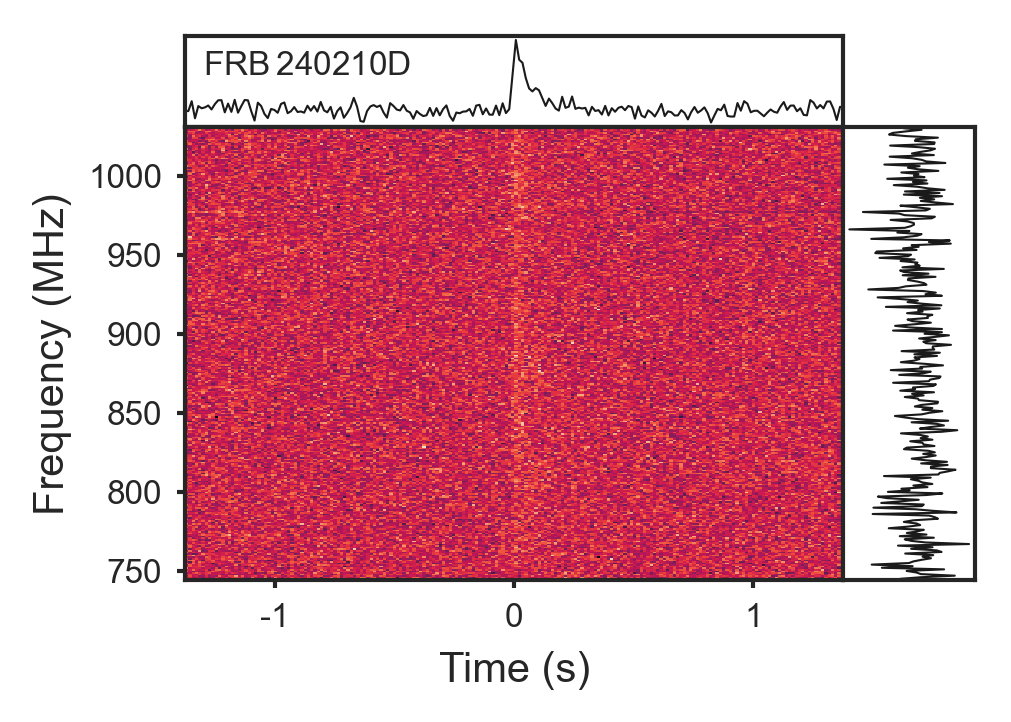}
    \includegraphics[width=0.49\linewidth]{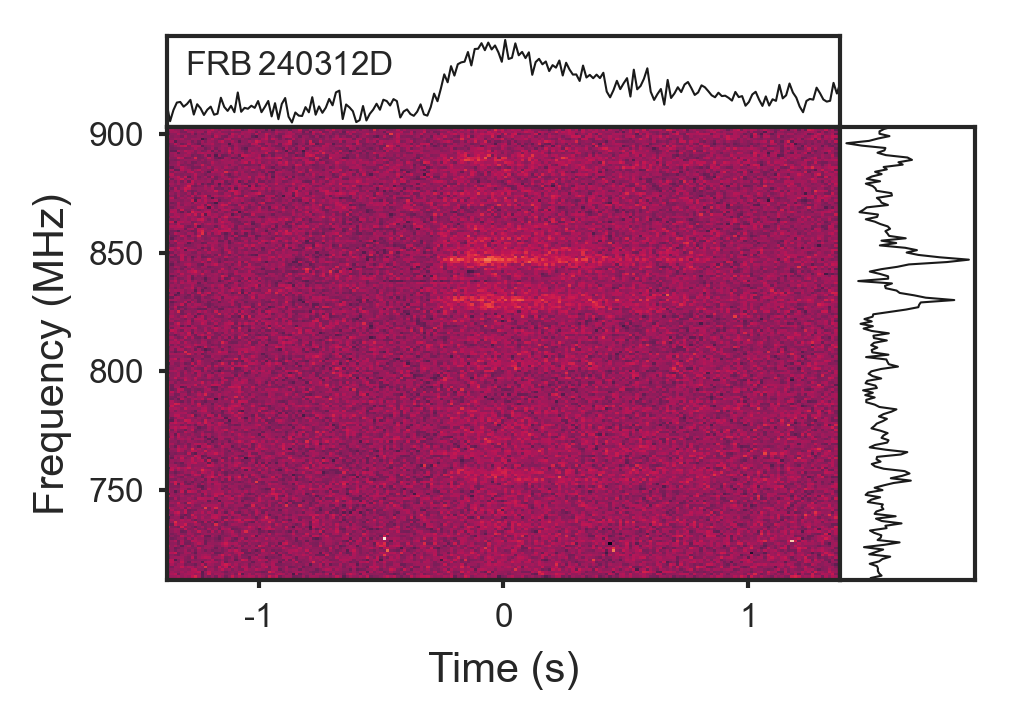}
    \caption{Waterfall plots of the two presented FRBs. The top panel shows the band integrated time series and the right panel shows the spectrum integrated over the on-burst region.}
    \label{fig:waterfaller}
\end{figure*}

\FRBo and \FRBt were discovered with ASKAP during the commissioning of CRACO.
Because these were commissioning observations, triggering of ASKAP's voltage buffers from the CRACO system was not yet enabled.
Instead, the intensity data were recorded at a time resolution of \SI{13.8}{\ms} for offline processing.
The recorded frequency band is divided into 240 channels of \SI{1}{\MHz} each.
The central frequency of the band changes depending on the specific science project being executed by ASKAP.
\FRBo was found in a CRAFT filler observation of G217+50 and \FRBt commensally with the FLASH survey. 
Since the recording only includes intensity data no polarisation information is available.

In what follows, we summarise the most important aspects of the detection pipeline and refer the reader to a more detailed description in \citet{Wang2025} and \ywinprep.
The recorded visibilities of the inner 24 antennas are used to make 256 by 256 pixel dirty images at 1024 trial DMs and 8 temporal widths, linearly up to 8 times the recorded time resolution.
Candidates from differential imaging are classified mainly by their distribution in the image plane, based on the observation that RFI is usually spread over the image while astronomical sources are concentrated at a specific point.
Astronomical source positions are compared to catalogues to identify pulsars and bright scintillating continuum sources.
The remaining candidates are inspected by eye and mainly include satellites, some other remaining RFI, spacial aliases of pulsars and scintillating continuum sources, new galactic radio transients, and FRBs.
The last two are differentiated by comparing their DM with Galactic electron density models.

We process the FRBs further for better localisation and generate a filterbank using the best location.
We form images and localise the FRBs within them using \tool{casa} \citep{CASATeam2022}; these images are astrometrically aligned using specially developed reference catalogues \citep{Jaini2025}.
More details will be presented by \ywinprep.

Both reported FRBs, shown in \cref{fig:waterfaller}, display visible scattering tails; additionally \FRBt shows scintillation. 

We begin the quantitative analysis by loading the filterbank intensity data from a time window of 2048 time samples around the burst time. We use \tool{sigpyproc}\footnote{\url{https://github.com/FRBs/sigpyproc3}} for this process, and also to dedisperse the bursts to approximate DMs of 387 and \SI{334}{\dmu}, respectively. An off-burst region is identified to calculate channel statistics, and the median-absolute-deviation, implemented in \tool{sigpyproc}, is used to mask bad channels. We normalise the data in each channel by subtracting the mean of the off-burst region and dividing by the standard deviation.

\subsection{Scattering}

\begin{figure}
    \centering
    \includegraphics[width=1\linewidth]{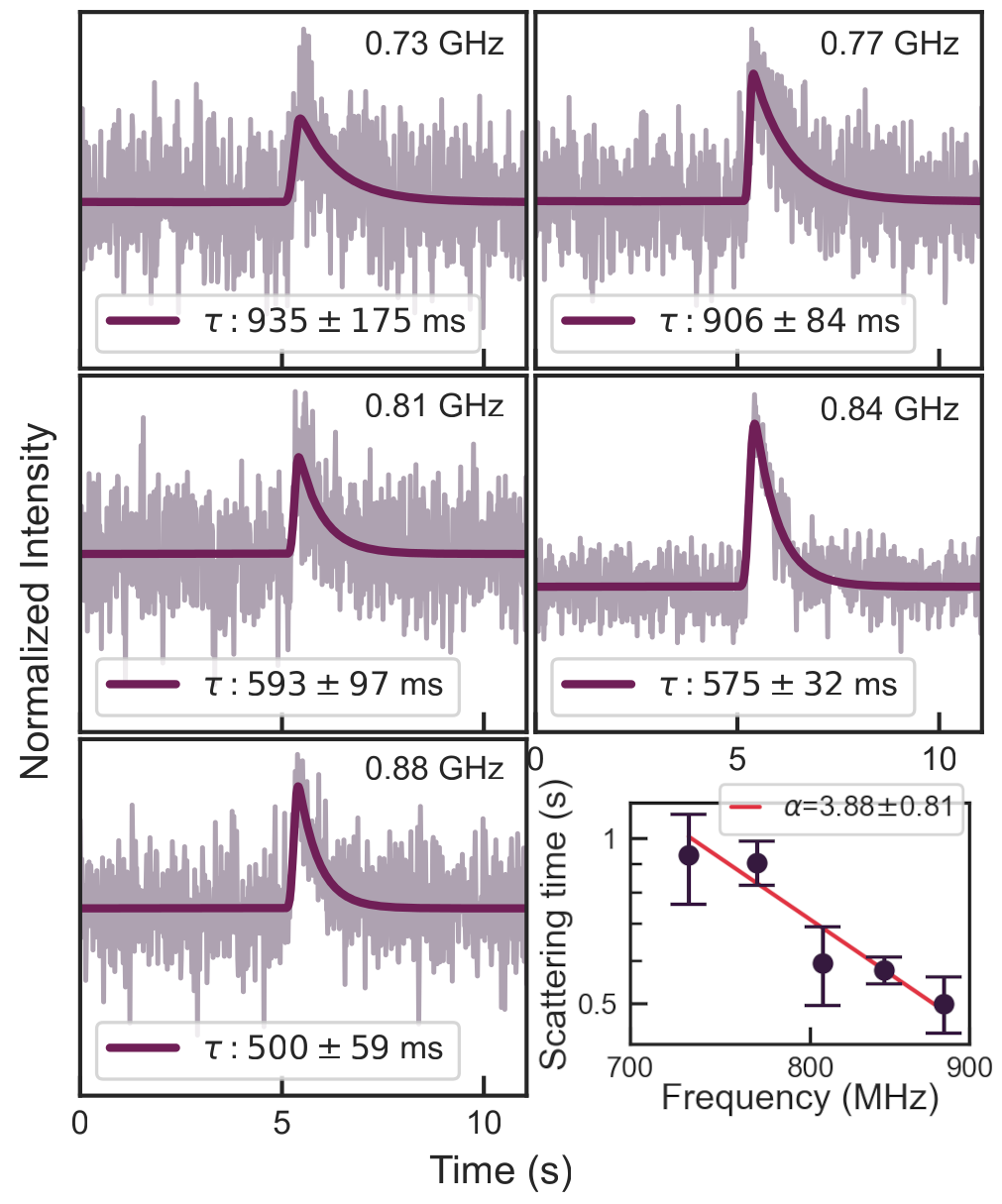}
    \caption{Fits to five subbands of equal bandwidth of \FRBt.}
    \label{fig:scattering}
\end{figure}

To measure the scattering time, we fit a single Gaussian convolved with an exponential scattering tail.
To confirm the presence of scattering, we perform the fit independently in several subbands.
For the fitting process, we select a smaller time window of 100 (800) samples around \FRBo (\FRBt).

We select a number of subbands and integrate over frequencies to obtain a time series for each subband. The number of subbands used is somewhat subjective: to include a DM in the fitting, at least 3 subbands are required, but the fit fails to converge when too many subbands are used.
In the end, we used four subbands for \FRBo and five subbands for \FRBt.
\Cref{fig:scattering} shows the subband time series of \FRBt along with the fitted functions.

The fits are done using \tool{scamp-i}\footnote{\url{https://github.com/pulsarise/SCAMP-I}} \citep{Oswald2021}, which works in the following way.
It employs \tool{emcee}\footnote{\url{https://emcee.readthedocs.io}} \citep{ForemanMackey2013} to independently perform an MCMC fit on each time series. The fit includes five parameters: the Gaussian amplitude, time, width, scattering time $\taus$, and a constant offset representing the baseline.
\tool{scamp-i} further uses the obtained time centres and the subband frequencies to fit for an improved DM using \tool{lmfit}.
This improved \DM is used for all following analyses.
Finally, a second MCMC is used to fit a two-parameter power law $\tau_\mr{subb} = \tau_\mr{ref} \times (\nu_\mr{subb}/\nu_\mr{ref})^\alpha$ between the scattering times $\tau_\mr{subb}$ and the subband central frequencies $\nu_\mr{subb}$; $\nu_\mr{ref}$ is the reference frequency taken to be the band centre and $\tau_\mr{ref}$ is the scattering time at $\nu_\mr{ref}$.
This fit is shown in the bottom right panel of \cref{fig:scattering} for \FRBt.

\subsection{Scintillation}

\begin{figure}
    \centering
    \includegraphics[width=1.\linewidth]{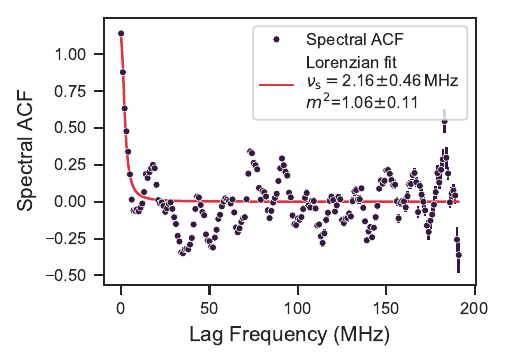}
    \caption{The ACF of \FRBt fitted by \cref{eq:Lorentzian}. Drawn error bars show the statistical uncertainties from noise.}
    \label{fig:scintillation}
\end{figure}

We measure the scintillation bandwidth of \FRBt as a fit parameter of a Lorentzian function, fitted to the autocorrelation function (ACF) of the spectrum \citep[see e.g.][]{Nimmo2025,Pradeep2025}.
To obtain the spectrum, we select a smaller time window by eye that contains signal-dominated times. The spectrum is computed by integration over this time window.
Our definition of the ACF differs from the other common definition in that we take the average of the intensity product instead of the sum \citep[eq.~(4.10) of][]{Pradeep2025}; this is needed to fit for the modulation index $m$.
We calculate one side of the ACF, exclude the zero-lag sample, and fit a Lorentzian function of the form
\begin{equation}
    f(\nusc)= \frac{m^2}{1 + (\delta\nu/\nusc)^2}\,, \label{eq:Lorentzian}
\end{equation}
where $\delta\nu$ is the frequency lag and $\nusc$ is the scintillation bandwidth.
The ACF of \FRBt along with the fit are shown in \cref{fig:scintillation}.
The plotted uncertainties of the ACF are also used in the fit weighting. They are calculated for the $i$-th ACF lag given $n$ frequency channels with signals $s_j$ and uncertainty $\sigma_j$ in the following way:
For $i < n/2$,
\begin{alignat}{1}
    \sigma_\mathrm{ACF}(i) = \frac{1}{n_\mathrm{valid}} \Bigg(&\sum_{j=0}^{i-1} \!\left(\sigma_j\, s_{j+i}\right)^2 + \sum_{j=i}^{n-i-1} \!\left(\sigma_j\left(s_{j+i} + s_{j-i}\right)\right)^2 \\ 
    & + \sum_{j=n-i}^{n-1} \!\left(\sigma_j\, s_{j-i}\right)^2 \Bigg)^{1/2}\,,
\end{alignat}
where $n_\mathrm{valid} = n - i$ for non-masked data or the count of unmasked overlapping pairs otherwise.
For $i \geq n/2$,
\begin{equation}
    \sigma_\mathrm{ACF}(i) = \frac{1}{n_\mathrm{valid}} \sqrt{ \sum_{j=0}^{n-i-1} \!\left(\sigma_j\, s_{j+i}\right)^2 + \sum_{j=i}^{n-1} \!\left(\sigma_j\, s_{j-i}\right)^2 }\,.
\end{equation}

We also tried two method variations: allowing for a constant offset in the fit, and excluding the burst spectrum by subtracting a smoothed spectrum.
The smoothed spectrum was obtained using a Gaussian with a width of 1/10th of the band.
The two results and the previous result all agree within their 68 per cent uncertainties so we use the first one.

\subsection{Energy}

The isotropic equivalent energy can be calculated through
\begin{equation}
    E = 4\pi D_L^2 F \Delta\nu/(1+z)\,,
\end{equation}
where $F$ is the fluence, $z$ is the redshift of the host galaxy, and $D_L$ the corresponding luminosity distance.

To calculate the FRBs' fluence, we use the dedispersed dynamic spectrum to compute the frequency averaged time series.
In case of \FRBo, we then subtract the mean of the off-burst region.
For \FRBt, we find that the baseline is not constant, so instead of assuming a constant baseline, we fit a line to the off-burst region (finding a slope of \SI{0.011}{\per\s}). After subtraction, we divide both bursts by the off-burst standard deviation. We then sum the signal over the time window that was also used for fitting the scattering time. The fluence is calculated using the radiometer equation:
\begin{equation}
    F = \mr{SNR}_\Sigma\times \frac{\mr{SEFD}}{\sqrt{N_\mr{ant}(N_\mr{ant}-1)}}\times\sqrt{\frac{\delta t}{2\times \Delta\nu}}\,,
\end{equation}
where $\mr{SNR}_\Sigma$ is the sum of \SNR{} over the time window, $\delta t=\SI{13.8}{\ms}$ is the sampling time, $\Delta\nu$ is the bandwidth, $N_\mr{ant}$ is the number of recorded antennas, and $\mr{SEFD}$ is the telescope's system equivalent flux density.

\subsection{Rates}

We aim to estimate the occurrence rate of long-duration FRBs based on the two reported events.
The real rate $R_\mr{tot}$ (of \si{\FRBs\per\sky\per\day}) above some threshold can be calculated from the observed rate $R$ and the fraction of the parameter space covered, i.e. the probability to detect a random FRB in the given experiment $P_\mr{obs}$:
\begin{equation}
    R_\mr{tot}=\frac{R}{P_\mr{obs}}\,.\label{eq:rtot}
\end{equation}
We chose this full treatment over others \citep[e.g.][]{Keane2015} because one of our sources was of longer duration than the largest search boxcar.

The observed rate can be inferred using the Poisson distribution from the number of detections $k=2$, the survey sky area $\Omega=\SI{1.21}{\deg\squared\per beam}\times\SI{36}{beams}$, and the time spent on sky $T=\SI{56}{\day}$, which we determined from recorded phase centre filterbank files.
The posterior probability for the rate assuming a flat prior for $R$ is 
\begin{equation}
    P(R\mid k)=\frac{(R\,\Omega\,T)^k\,\Omega\,T}{k!}\,e^{-R\,\Omega\,T}\,.\label{eq:rate}
\end{equation}

We estimate $P_\mr{obs}$ by sampling our parameter space in a grid and integrating over it.
First, we need to make assumptions for the FRB population, then we can apply survey responses.
For the FRB fluence distribution, we assume a power-law $P(F)\propto F^{-3/2}$. 
We further assume a uniform probability for the position in the beam and a log-uniform distribution for the burst widths.

The probability to detect a burst depends only on the \SNR, which in the case of CRACO depends mainly on three factors.
\begin{enumerate}
    \item The mismatch between a Gaussian burst with width $w_t$ and the closest boxcar width $w_\mr{bc}$:
\begin{equation}
    \eta_{w_t}(w_t, w_\mr{bc})=\mr{erf}\left(\frac{w_\mr{bc}}{w_t}\right)\sqrt{\frac{w_t}{w_\mr{bc}}}\,,
\end{equation}
where $\mr{erf}$ is the error function, and we defined $w_t=2\sqrt{2}\sigma=2.83\sigma$ the optimal boxcar width of a Gaussian burst.
\item The beam response at the burst angular distance from the beam centre $\theta$, $\eta_\mr{beam}(\theta)$, which we assume to be Gaussian with a full width at half maximum of \SI{1.8}{\degree}.
\item The position-dependent loss in \SNR due to the way the visibilities are gridded \citep{Wang2025},
\begin{equation}
    \eta_\mr{grid}(l, m)=\frac{\sin(\pi\,l/\Delta l)}{\pi\,l/\Delta l}\,\frac{\sin(\pi\,m/\Delta m)}{\pi\,m/\Delta m}\,,
\end{equation}
where $l$ and $m$ are the pixel coordinates, and $\Delta l=\Delta m=\SI{1.1}{\degree}$ are the side lengths of the grid.
The \SNR of a burst is then given by
\begin{equation}
    \SNR=\sqrt{\frac{2\,\Delta\nu}{w_t}}\frac{F}{\mr{SEFD}}\,\eta_{w_t}(w_t, w_\mr{bc})\,\eta_\mr{beam}(\theta)\,\eta_\mr{grid}(l, m)\,,
\end{equation}
where $\Delta\nu$ is the searched bandwidth, and $w_t$ the burst width.

\end{enumerate}

Integrating over all possible observed \SNR that would result in a detection, the probability to detect a burst is
\begin{alignat}{1}
    P(\SNR_\mr{th},\Theta)&=\int_{\SNR_\mr{th}}^\infty\mathcal{G}(\rho\mid\SNR,1)\,\mr{d}\rho\\
    &=\frac{1}{2}\left(1+\mr{erf}\left(\frac{\SNR(\Theta)-\SNR_\mr{th}}{\sqrt{2}}\right)\right)\,,
\end{alignat}
where $\mathcal{G}$ denotes the normal distribution and $\Theta=\{F,w_t,l,m\}$.
The total probability following from the integral
\begin{equation}
    P_\mr{obs}=\iiiint P(\SNR_\mr{th},\Theta)p(F)p(w_t)\,\mr{d}F\,\mr{d}w_t\,\mr{d}l\,\mr{d}m\,. \label{eq:Pobs}
\end{equation}
With this, we have all the ingredients for \cref{eq:rtot}.

\section{Results}
\label{sec:results}

\begin{table*}
    \centering
    \caption{Key values for the two FRBs.}
    \begin{tabular}{lrr}
        \toprule
        	&	\FRBo	&	\FRBt	\\
        \midrule					
        search DM (\si{\dmu})	&	396	&	330	\\
        RA	&	03:40:55.979$\pm$2.731 arcsec	&	03:26:29.613$\pm$1.1 arcsec	\\
        RA (deg)	&	55.23325(76)	&	51.62339(31)	\\
        Dec	&	-41:51:43.6$\pm$1.3 arcsec	&	-54:36:03.5$\pm$1.3 arcsec	\\
        $l$ (deg)	&	247.282207	&	268.0116392	\\
        $b$ (deg)	&	-52.5575921	&	-50.684932	\\
        	&		&		\\
        Nr. Subbands	&	4	&	5	\\
        DM (\si{\dmu})	&	\num{385.7\pm1.3}	&	\num{333.8\pm3.8}	\\
        $\nu_\mathrm{ref}$ (MHz)	&	888	&	807	\\
        $\taus[\mr{ref}]$ (ms)	&	68.0$\pm$6.6	&	688$\pm$31	\\
        $\taus[,\SI{1}{\GHz}]$ (ms)	&	34.0$\pm$6.1	&	300$\pm$48	\\
        $\taus[,\SI{1}{\GHz},\mr{rest}]$ (ms)	&	79$\pm$14	&	347$\pm$56	\\
        $\alpha$	&	5.9$\pm$1.4	&	3.883$\pm$0.031	\\
        $\taus$ (ms) (NE2025)	&	\num{5.6e-05}	&	\num{6.3e-05}	\\
        $\taus$ (ms) (YMW16)	&	\num{1.3e-04}	&	\num{1.3e-04}	\\
        $\nusc$ (MHz)	&	$< 1$ or $>60$	&	\num{1.32\pm0.33}	\\
        $m$	&		&	$1.03\pm0.05$	\\
        $\Dhs D_\mr{MW}$ (\si{\kpc\squared})	&		&	$\lesssim0.0083 \pm 0.0018$	\\
        \DMMW (ne2025)	&	33	&	34	\\
        \DMMW (YMW16)	&	26	&	27	\\
        \DMX (\si{\dmu})	&	$15^{+3}_{-4}$	&	$21^{+15}_{-8}$	\\
        \DMd (\si{\dmu})	&	$223^{+109}_{-40}$	&	$28^{+22}_{-8}$	\\
        \DMhalo (\si{\dmu})	&	$40\pm10$	&	$40\pm10$	\\
        \DMh (\si{\dmu})	&	$81_{-81}^{+42}$	&	$214^{+16}_{-29}$	\\
        $\DM_\mr{host,rest}$ (\si{\dmu})	&	$107_{-107}^{+55}$	&	$225^{+17}_{-30}$	\\
        Fluence (\si{\Jy\ms})	&	\num{28.6\pm6.1}	&	$276\pm57$	\\
        Energy (\si{\erg})	&	\num{4.8\pm1.0 e38}	&	\num{29.4\pm6.1 e38}	\\
        Specific Energy (\si{\erg\per\Hz})	&	\num{16.9\pm3.6 e29}	&	\num{16.3\pm3.4 e30}	\\
        \midrule					
        \multicolumn{3}{c}{Host galaxy}					\\
        \midrule					
        					
        Names	&		&	2MASX J03262952-5435569	\\
        	&		&	GALEX J032629.4-543557	\\
        PATH probability	&	0.89	&	0.99	\\
        $z$	&	$0.3247\pm0.0003$	&	\num{0.04986(2)}	\\
        Distance (\si{\Mpc})	&	1756	&	230	\\
        Stellar Mass (\si{\Msol})	&		&	$\num{2\pm 1e10}$	\\
        SFR (\si{\Msol\per\yr})	&		&	2$\pm$0.06	\\
        \bottomrule
        \multicolumn{3}{l}{\footnotesize{Quantities not defined in the text: $\DM_\mr{host,rest}=\DMh(1+z)$}}
    \end{tabular}
    \label{tab:results}
\end{table*}

All measured properties of the FRBs are summarised in \cref{tab:results}.

\subsection{\FRBt}

We measure a scintillation bandwidth of $\nusc=1.32\pm\SI{0.33}{MHz}$ at \SI{808}{\MHz}.
This is compatible with the values predicted for the Milky Way from electron models $\nusc[\mr{,MW}]\approx 1/2\pi\taus[\mr{,NE2001}]=\SI{2.5}{MHz}$ or \SI{1.2}{MHz} for the YMW16 model.
On the other hand, \nusc cannot originate from the same screen as the observed scattering because it would only cause a broadening of $\taus=1/2\pi\nusc=\SI{0.12}{\us}$.
Together with the high Galactic latitude, this strongly suggests at least two scattering screens where the scintillation comes from a screen in the Milky Way and the scatter broadening comes from an extragalactic screen \citep[see also][]{Masui2015}.

Since the source is scintillating in the Milky Way, the scattered images angular size must be smaller than the resolution of the Milky Way screen.
This argument has, e.g., been used historically to constrain the angular size of quasars, scintillating in the solar corona \citep{Hewish1964}.
Following \citet{Pradeep2025}, this gives us a constraint on the following distance product:
\begin{equation}
    \Dhs D_\mr{MW}\lesssim\frac{(1+z)\Ds^2}{8\pi\nu^2}\frac{\nusc}{m\taus}=0.0083 \pm \SI{0.0018}{\kilo\pc\squared}\,, \label{eq:dist_prod}
\end{equation}
where $\Dhs$ is the distance between source and host galaxy screen, $D_\mr{MW}$ is the distance between the observer and the screen in the Milky Way, and $\nu$ is the observing frequency \citep{Sammons2023}.
Because we did not find a significant deviation of $m$ from 1, we set $m=1$.
We note that \cref{eq:dist_prod} assumes two-dimensional screens.
In the case of two one-dimensional screens it is only true when their sky projection is parallel with the constraint getting weaker towards a perpendicular orientation \citep[see e.g. section 7.3 of][]{Pradeep2025}.

\subsection{\FRBo}

This FRB shows no scintillation within the sensitive scintillation bandwidth ranges limited by bandwidth and channel width.
The scattering times predicted for this sightline by the NE2001 and YMW16 models are $<\SI{1}{\us}$.
The screen causing the observed scattering is therefore likely to be extragalactic and could be either in the host galaxy or in an intervening galaxy or large scale structure.

\subsection{Rate of long duration FRBs}

Because the parameter space of FRBs beyond \SI{100}{\ms} duration is not well probed, it is particularly interesting to calculate the rate.
We do so based on the two detections reported here that were the only non-repeaters found with a search boxcar of 4 times the time resolution or higher in the considered time frame.
One more burst, \FRBYrep was detected with a boxcar width of 4, but it repeated later and will be discussed in \ywinprep.
During commissioning, the system performance can change causing some uncertainty in the sensitivity.
Yet, judging from the detection rates, the performance was stable between MJD~60304 and approximately MJD~60408.
During this time, CRACO searched for 56 days.

Because both FRBs were detected above a boxcar width of $4\times\SI{13.8}{\ms}=\SI{55.2}{\ms}$, we estimate the rate of bursts with durations between \SI{55.2}{\ms} and \SI{1}{\s} and above a fluence of \SI{9}{\Jy\ms}, which is the detection threshold at the lowest duration.
Above these thresholds, we find that we should detect $P_\mr{obs}=0.16$ (see \cref{eq:Pobs}) of the FRBs.

From \cref{eq:rate}, we obtain the observed rate $R=34^{+74}_{-28}\,\si{\FRBs\per \sky\per \day}$ (mode and 95 per cent highest density interval).
This yields a total rate of $R_\mr{tot}=210^{+460}_{-180}\,\si{\FRBs\per \sky\per \day}$.

\section{Multiwavelength follow-up}
\label{sec:followup}

\subsection{\FRBt}
\subsubsection{Host galaxy identification and follow-up observations}

\begin{figure*}
    \centering
    \includegraphics[width=0.49\linewidth]{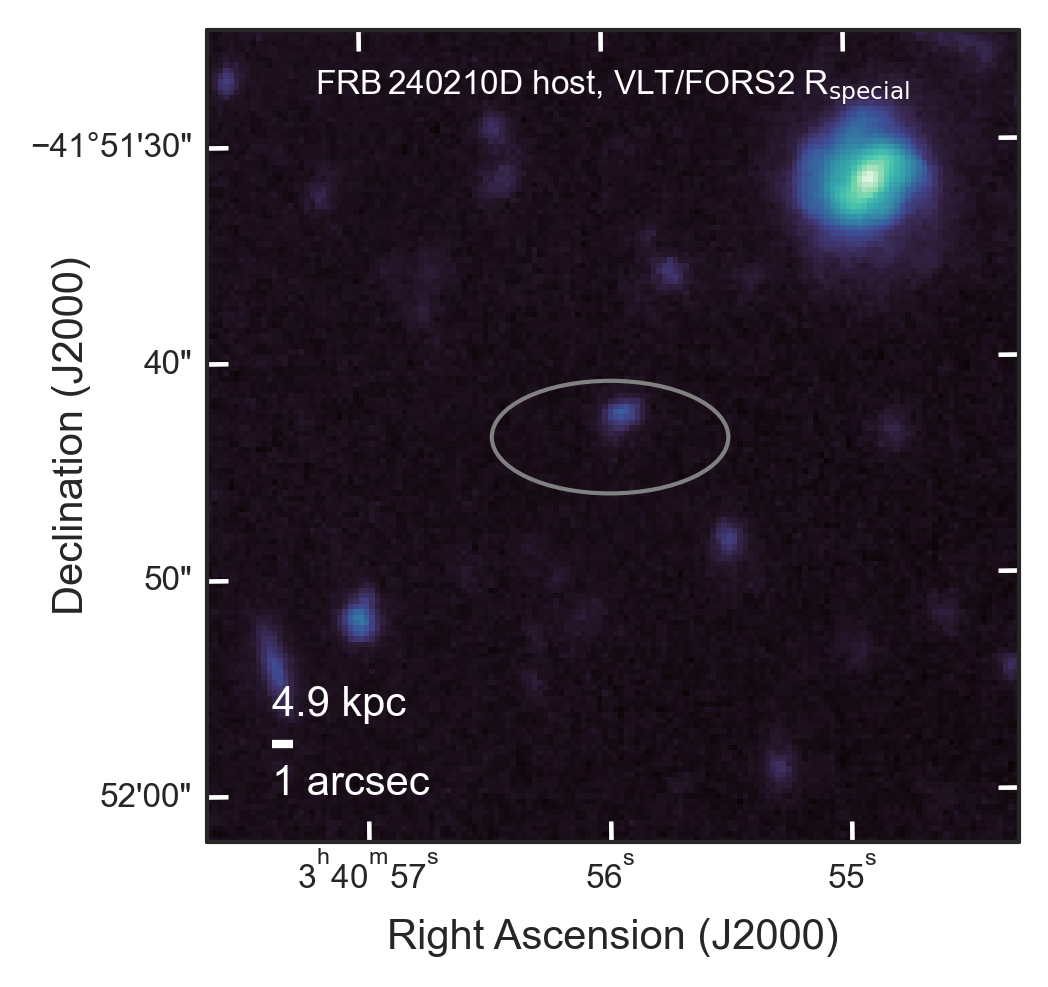}
    \includegraphics[width=0.49\linewidth]{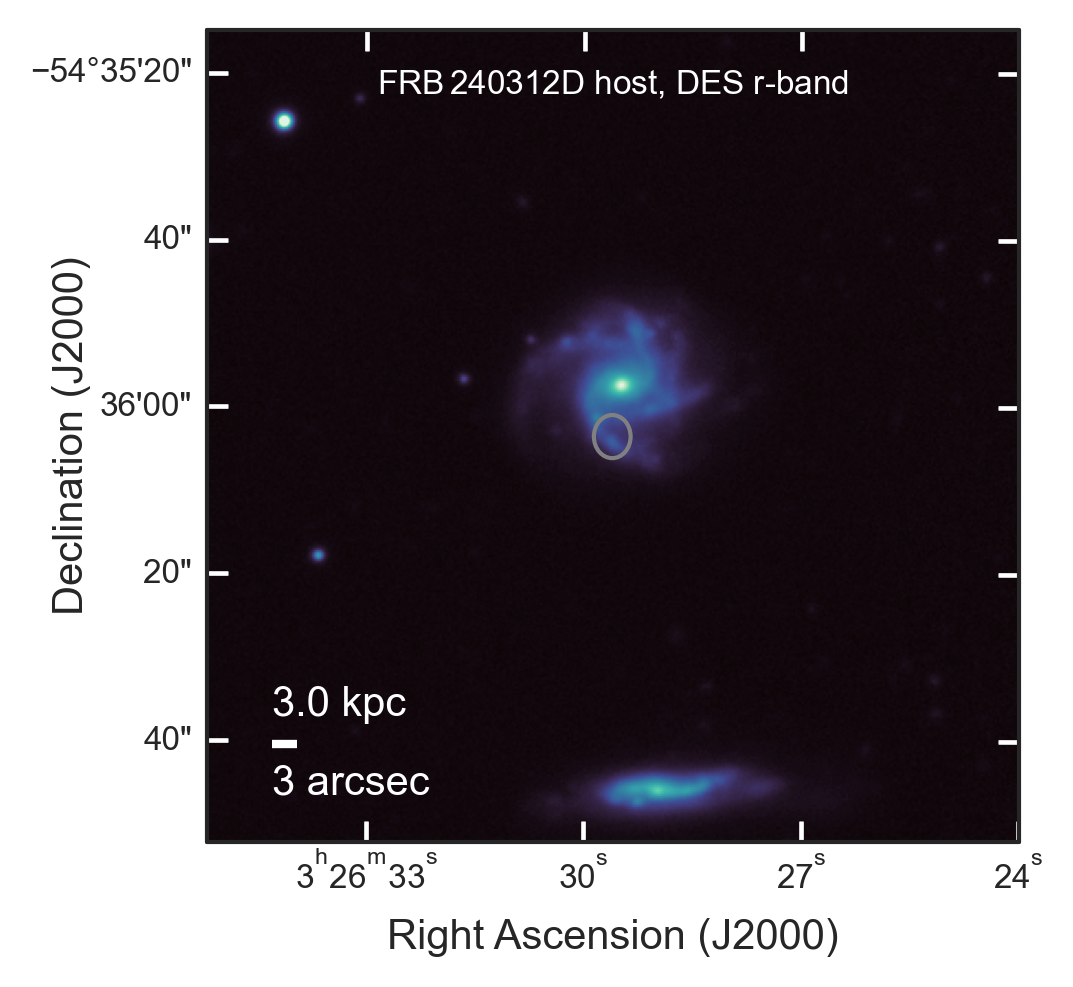}
    \caption{Host Galaxies of \FRBo (left, obtained under \texttt{Program ID 108.21ZF}, \texttt{PI Shannon}) and \FRBt \citep[right, with data from][]{Abbott2021c}. The flux axis is stretched using the inverse hyperbolic sine function. The grey ellipse shows the 2$\sigma$ localisation corresponding to the sky area containing the FRB with 90 per cent probability.}
    \label{fig:HGs}
\end{figure*}

We first searched for potential host galaxies in existing catalogues using the Data Aggregation Service \citep{Miszalski2022}. For \FRBt, we identified a host galaxy with a catalogued redshift in the 6dF galaxy survey \citep{Jones2009} and several other surveys; an image of it is shown in \cref{fig:HGs}.

We also obtained additional spectra of the \FRBt host galaxy and the edge-on spiral galaxy south of it.
We used the Goodman Spectrograph on the 4-m Southern Astrophysical Research Telescope \citep[SOAR]{Clemens2004} under \texttt{Program ID SOAR-2024A-002} (\texttt{PI Gordon}).
The data have previously been published by \citet{Gordon2025}, who also described the data reduction and the detected lines.
We confirm the UKST/6dF spec-$z$ of the host to be $z=0.0495$ and we measure the edge-on neighbour, 2MASS J03262894-5436459, to be at a slightly higher redshift of $z=0.0521$.
We will discuss an ambiguous group membership below.

In order to characterise the interstellar medium (ISM) of the \FRBt{} host galaxy and search for gas associated with or affected by the FRB source, we use spectroscopic data from Multi Unit Spectroscopic Explorer \citep[MUSE;][]{bacon2010} mounted on UT4 of the Very Large Telescope (VLT). MUSE observations (\texttt{Progarm ID: 114.27C4.001}, \texttt{PI: N.Tejos}) were conducted in the `wide-field adaptive optics' mode covering $1\arcmin \times 1\arcmin$ FOV with spacial sampling of $0.2\arcsec~{\rm pix}^{-1}$ and wavelength coverage of $\SIrange{4700}{9300}{\angstrom}$. The field was integrated for a total of $8\times 600$s with seeing of $0.8\arcsec$. The data were reduced using the standard ESO MUSE pipeline \citep{Weilbacher2020} with standard setup and parameters.

To analyse the data cube, we used the \tool{mpdaf} package\footnote{\url{https://github.com/musevlt/mpdaf}} \citep{Piqueras2017} and specifically \tool{pyplatefit}\footnote{\url{https://github.com/musevlt/pyplatefit}} to perform the spectral line fits separately in each spaxel.
We provide the measured galaxy redshift as an initial redshift and fit for all the default lines, which are H$\alpha$, H$\beta$, \OIIIa, \OIIIb, \NIIa, \NIIb, \SIIa, \SIIb, \OI, \HeI, and \ArIII.
The lines are split into two families: Balmer and all others.
Within the families, the velocity offset and velocity dispersion are shared and jointly fit.
We employ \tool{PyNeb}\footnote{\url{https://pypi.org/project/pyneb/}} \citep{Luridiana2015} to measure the gas density from the \SIIa{}$/$\SIIb ratio.
We repeat the same analyses after rebinning in the spacial dimensions by a factor of 2 to increase the $S/N$ of lines.

\subsubsection{Potential group membership}
\begin{table}
    \centering
    \caption{Coordinates of the potential group members identified in H\,\textsc{i} \citep{Roxburgh2025}. Measurements apart from the first two lines stem from the H\,\textsc{i} data.}
    \begin{tabular}{rrSrl}
        \toprule
        RA (deg)	&	Dec (deg)	&	$z$	&	Offset (kpc)$^\text{(a)}$	&	Comment	\\
        \midrule									
        51.6229	&	-54.5993	&	0.04986	&	159	&	FRB host	\\
        51.6206	&	-54.6128	&	0.0521	&	192	&	Neighbour	\\
        51.5235	&	-54.6387	&	0.04783	&	791	&		\\
        51.4401	&	-54.6158	&	0.04735	&	1290	&		\\
        51.7784	&	-54.5009	&	0.0466	&	871	&		\\
        51.8177	&	-54.5175	&	0.04672	&	1088	&		\\
        51.6378	&	-54.5793	&	0.04639	&	61	&		\\
        51.7356	&	-54.6181	&	0.04641	&	563	&		\\
        \bottomrule
        \multicolumn{5}{l}{\footnotesize{$^\text{(a)}$The projected offset from the (non-weighted) mean.}}
    \end{tabular}
    \label{tab:group}
\end{table}

The southern edge on neighbour galaxy (see \cref{fig:HGs}) brings up the question of group embedment.
If the two galaxies are part of the same group, the difference in redshift would imply a radial velocity difference of \SI{780}{\km\per\s}, largely exceeding typical velocities in galaxy groups \citep[see e.g.][]{Tully2015}.
On the other hand, the chance coincidence of finding two unrelated galaxies so close to each other seems low as well.

Furthermore, \citet{Roxburgh2025} found that the FRB host is likely in an H\,\textsc{i}-rich group environment with up to eight H\,\textsc{i} detections at similar redshifts.
We list the coordinates and redshifts of the potential group members in \cref{tab:group}.
In the following, we calculate the virial mass for different configurations using equation (2) from \citet{Tully2015}.

The edge-on neighbour galaxy cannot be in a group at the FRB distance for several reasons.
In accordance with the large radial velocity difference, we find a virial mass of $M_\mr{vir}\gtrsim\SI{2e14}{\Msol}$ when we include it in the group.
On the other hand, we do not find any cluster or massive group detected in eROSITA \citep[extended ROentgen Survey with an Imaging Telescope Array;][]{Bulbul2024} suggesting a mass of $M_\mr{vir}<\SI{e13}{\Msol}$.
Furthermore, there is no sign of an interaction in the H\,\textsc{i} maps \citep{Roxburgh2025}.
Even though there is a hint of a tidal disturbance of the FRB host, it is at the opposite side of the edge-on galaxy.
Given the small projected distance, the disturbance would have to be much more prominent \citep{ArgudoFernandez2013}.
In consequence, the larger redshift of the edge-on galaxy must largely be from the Hubble flow, which places it firmly in the background.

The FRB host and the remaining six galaxies are also unlikely to be in a group.
Excluding the background galaxy yields $M_\mr{vir}\sim\SI{1.5e+14}{\Msol}$, still too large for an X-Ray non-detection.
Further excluding the host yields a possible scenario with $M_\mr{vir}\sim\SI{4e+13}{\Msol}$.
However, this implies a very large geometric radius of \SI{1.9}{\Mpc}, which is as well incompatible with the size of a group.
We note that if it nevertheless was a foreground group, it would impose a DM of $49^{+47}_{-22}\,\si{\dmu}$.

Given the uncertain nature of the structure that the galaxies are part of and the large distances, we do not include it in our DM budget.
They might indicate the presence of a cosmic filament, but such an identification is beyond the scope of this paper.

\subsubsection{Host galaxy}
The image of the host galaxy is shown in \cref{fig:HGs} with the 90 per cent probability localisation  of the FRB.
The host of \FRBt is a spiral galaxy of morphological type Sc \citep{Moustakas2023}.
Based on the line fits to the MUSE data, it is located at $z=\num{0.04986(2)}$, corresponding to a luminosity distance of \SI{230}{\Mpc}.
Its half-light radius is \SI{6.63}{\kpc} \citep[see table 3 of][]{Gordon2025}.
Further catalogued properties are summarised in \cref{tab:results}.

We measure the global galaxy properties as follows.
We adopt the galaxy's mass estimate of \SI{2\pm 1e10}{\Msol} from the GLADE catalogue \citep{Dalya2022} and infer the star formation rate (SFR) in two independent ways: from its UV emission following \citet{Mannings2021} and from the H$\alpha$ flux.
We use the catalogued near UV flux from GALEX \citep{Bianchi2014} and the luminosity-SFR relation from \citet{Kennicutt2012} to calculate an SFR of $\mathrm{SFR}=\SI{2\pm0.06}{\Msol\per\yr}$.
The H$\alpha$ flux measurement throughout the galaxy in the MUSE data allows for an independent measurement.
Summing over all H$\alpha$ full resolution spaxels with $\SNR>3$, we obtain an H$\alpha$ flux of \SI{5281\pm 6e-17}{\erg\per\cm\squared\per\s}. This yields \citep{Kennicutt1998}
\begin{equation}
    \mr{SFR}=\num{7.9e-42}\, \frac{L(\mr{H}\alpha)}{\si{\erg\per\s}}\, \si{\Msol \per\yr} = \SI{2.622(3)}{\Msol \per\yr} \,.
\end{equation}

We measure the inclination angle from the data of the Dark Energy Survey \citep[DES;][]{Abbott2021c} using \tool{galfit} \citep{Peng2002} to be \SI{20.7\pm0.3}{\deg}, i.e., the galaxy is almost face on.
Here, the uncertainty only accounts for the uncertainty in the fit and does not include deviations of the galaxy shape from the model.
More details on the modelling process will be presented by Marnoch et al.\ (in prep.).

The FRB is located close to or in a spiral arm.
\citet{Gordon2025} established a statistical association with the spiral arm which is preferred over an ellipsoidal and spherical model with a Bayesian information criterion that is higher by 3.09 and 3.14, respectively.
The galactocentric offset is\footnote{The distance of \FRBt is coincidentally about $3600\times180/\pi\,\si{\kpc}$, such that $\SI{1}{\arcsecletter} \,\hat{=}\,\SI{1}{\kpc} $.} \SI{6.4\pm1.1}{\kpc} and hence inconsistent with the FRB being at the galaxy centre at about 5.8$\sigma$.

\subsubsection{Integral field spectroscopy}
\begin{figure*}
    \centering
    \includegraphics[width=\linewidth]{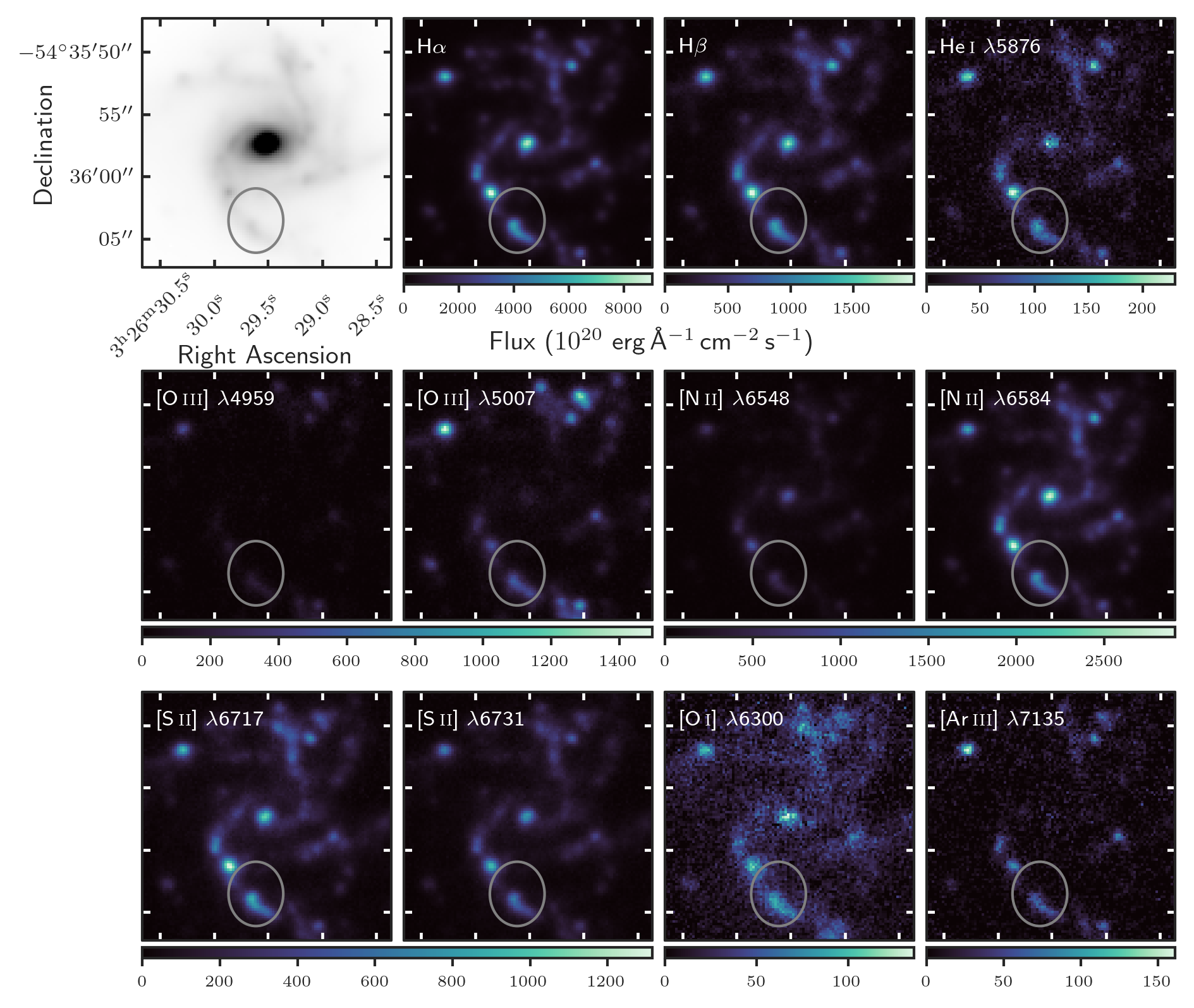}
    \caption{Maps of the emission lines we found in our VLT/MUSE observations of the \FRBt host. The grey ellipse shows again the FRB's 90 per cent probability localisation.}
    \label{fig:MUSE}
\end{figure*}

\begin{figure}
    \centering
    \includegraphics[width=\linewidth]{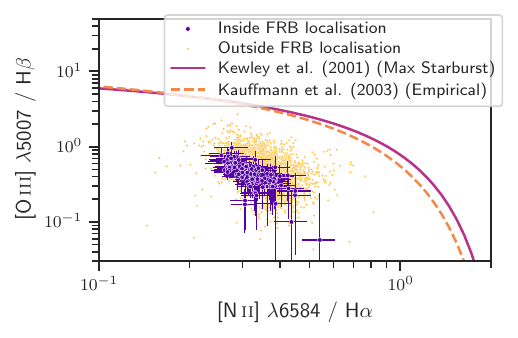}
    \caption{BPT diagram showing the line ratios that indicate the primary ionisation source for the rebinned spaxels in the \FRBt host. Lines show the laws separating radiation driven from AGN driven ionisation \citep[taken from][]{Kewley2001,Kauffmann2003}. Purple dots show the line ratios within the FRB 90 per cent localisation region with uncertainties, whereas the yellow ones are from the remainder of the galaxy.}
    \label{fig:BPT}
\end{figure}

\begin{figure*}
    \centering
    \includegraphics[width=\linewidth]{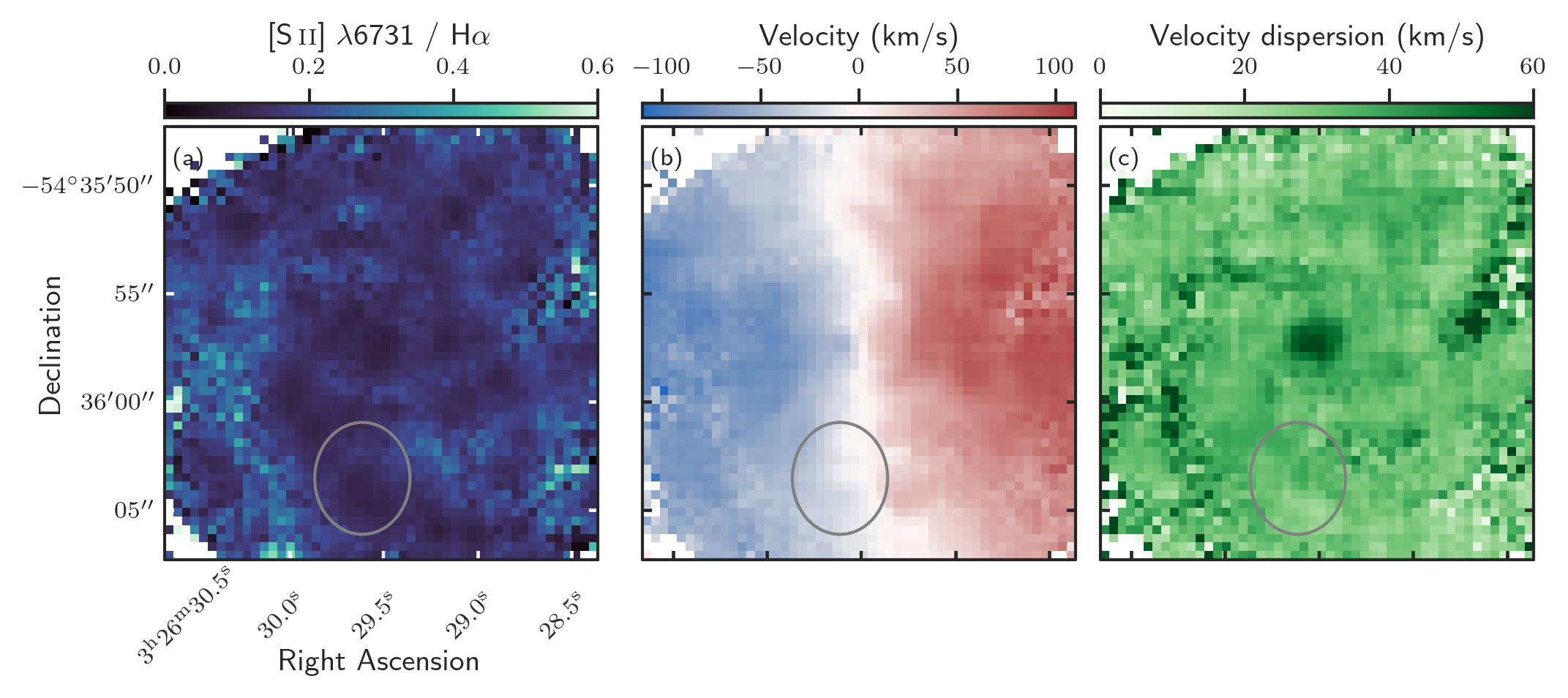}
    \caption{Line ratios and galaxy kinematics. Panel (a) shows an example line ratio map; this ratio is used to distinguish \HII regions from SNRs. Panel (b) and (c) show the velocity and velocity dispersion of the Balmer lines. Spaxels  have been masked, where the H$\alpha$ $\SNR<3$ (panel (a)) or the \SNR sum of H$\alpha$ and H$\beta$ $<3$ (panels (b) and (c)).}
    \label{fig:kinematics}
\end{figure*}

Within the MUSE data, emission lines of the gas in the galaxy are of primary interest.
For instance, the H$\alpha$ flux is a direct tracer of star formation.
To analyse the state of the gas, line ratios are of special importance.
Through these one can distinguish between gas that is photoionized, shock ionized gas in super nova remnants (SNRs), and gas ionized by an active galactic nucleus.
For example, a ratio of \SIIb{}$/$H$\alpha>0.4$ indicates shock ionized gas and has traditionally been used to identify SNRs \citep{Mathewson1973}.
The ratio of \SIIa{}$/$\SIIb is a tracer of the gas density.
Line redshifts and line widths provide information about the kinematics of the galaxy and its gas.

We detect all the major emission lines that we searched for throughout large parts of the galaxy.
\Cref{fig:MUSE} shows maps of the different line fluxes.
The maps show one knot close to the centre of the FRB extending southwards along the spiral arm.
A second bright knot at the northeast is at a branching point, where a spiral arm grows out of the main arm towards the west.
This knot itself appears unresolved but has weaker extensions towards the two arms.

Qualitatively comparing the two knots around the FRB with other features in the galaxy, none of the maps shows very striking differences.
The gas in the galaxy's centre is almost vanishing in [O\,\textsc{iii}] implying a higher ionisation state.
Knots in the north of the galaxy show elevated [O\,\textsc{iii}] and [Ar\,\textsc{iii}].
However, no such obvious behaviour is visible at the FRB location.

Line ratios are the primary tool to distinguish regions of star formation from active galactic nuclei and SNRs.
We use the rebinned maps for obtaining more precise line ratios in the following.
\Cref{fig:BPT} shows the Baldwin, Phillips and Terlevich (BPT) diagram \citep{Baldwin1981} of the spaxels in the host galaxy, with the ones inside the 2$\sigma$ region highlighted.
The line ratios within the FRB localisation ellipse show clear star formation driven emission lines.
And we note that also no other regions in the galaxy -- particularly the centre -- are outside the region indicating ionisation from star formation.

For the identification of SNRs, [S\,\textsc{ii}]/H\,$\alpha$ is classically used \citep{Dodorico1978,Kopsacheili2024}.
A map of the ratio is shown in panel (a) of \cref{fig:kinematics}.
Overall the dense regions in the galaxy are all well below a ratio of 0.4.
We measure \SIIb$/\;\text{H}\alpha=0.11\pm 0.01$ for both, the knot at the centre of the FRB position and the one in the northeast.
Beyond the classic line ratio, the clearest diagnostic suggested by \citet{Kopsacheili2020} is the \OI/H$\alpha$ ratio with the condition for an SNR being \OI$/$H$\alpha>0.017$.
We measure $\text{\OI}/\text{H}\alpha=0.015\pm0.007$ at the centre of the FRB central knot and $0.013\pm0.007$ in the northeastern knot.

The spacial resolution of the observations limits our ability to detect smaller objects like SNRs or pulsar wind nebulae (PWNe).
For instance, a \SI{1000}{\yr} old SNR has a radius of about \SI{10}{\pc}.
Our resolution corresponds to \SI{0.8}{\kpc} at the galaxy distance and hence the sky area with the SNR will only contribute $(1/40)^2=0.000625$ of a spaxel's integrated sky area.
Therefore, while the lines in the knots around the FRB are clearly dominated by ionisation from young stars, it is difficult to exclude the presence of any SNRs and even more so of PWNe, with typical sizes of 0.1 to \SI{1}{\pc}.

We constrain the gas density from the ratio [S\,\textsc{ii}]$\lambda6716/\lambda6731$.
Over most of the galaxy and also at the FRB location, the ratio is close to 1.4.
However, for ratios slightly above 1.45, the density diverges to 0 and differences from the assumed temperatures become more important \citep[see fig. 16 of][]{Kopsacheili2024}.
Because the uncertainties of the ratios are all above 0.2, we can only use it as an upper limit.
Taking the ratio of the average flux inside the localisation region we get a ratio of \num{1.43\pm0.03}, which allows us to set an upper limit of \SI{80}{\per\cm\cubed}.
At the given resolution, the line ratio is dominated by gas at typical ISM conditions.

The velocity offset of a spaxel with respect to the input redshift and the velocity dispersion is returned by the fitting routine for every spaxel and separately for the two line families.
The velocity field of the Balmer lines is shown in \cref{fig:kinematics} and shows a clear east-west gradient.
It agrees with the inner part of the H\,\textsc{i} velocity field shown in fig.\ 2 of \citet{Roxburgh2025}.
Assuming a trailing spiral pattern, as observed in the overwhelming majority of spiral galaxies \citep[e.g.][]{Binney2009}, we can infer that the angular momentum vector points away from the observer.
Thus, the southern half of the galaxy that includes the FRB source is the far side of the galaxy.
This could be included in future \DMh studies, but given the low host inclination, it is not included in our DM budget here.

The dispersion, which is measured from the line widths, is also shown in \cref{fig:kinematics} for the Balmer lines.
It is generally around \SI{32}{\km\per\s} throughout the galaxy, with higher values in the centre.
The H\,\textsc{ii} regions around the FRB position do not show any measurable enhancement at the given resolution.

\subsubsection{Intervening galaxies}
\label{sec:frbtinterv}

We searched for intervening galaxies in a cone of \SI{10}{\degree} around the FRB position.
For this, we used \tool{VizieR} to query the GLADE+ catalogue \citep{Dalya2022} for $z<0.05$ galaxies.
We use the available redshifts to calculate the distance to each galaxy, and subsequently the impact parameter $b$ from the angular offset.
We inspect all galaxies with an impact parameter and impact parameter uncertainty less than \SI{500}{\kpc}.
The three galaxies that pass this criterion are NGC~1311 ($b=\SI{230}{\kpc}$, GLADE+1328907), IC1959 ($b=\SI{420}{\kpc}$, GLADE+1328936), and ESO154-023 ($b=\SI{430}{\kpc}$, GLADE+1328865);
their masses from the literature are \SI{e12}{\Msol} \citep{Marino2010}, \SI{6e+8}{\Msol} \citep[baryon mass from][]{Kourkchi2022}, and \SI{e9}{\Msol} \citep[stellar plus H\,\textsc{i} mass from][]{Georgiev2010}, respectively.
Furthermore, NGC~1311 and IC1959 are part of the galaxy group LGG~93.

We estimate their imposed DM using the modified Navarro–Frenk–White profile of \citet{Prochaska2019} with concentration parameter $c=7.7$, model parameters $\alpha=2$ and $y_0=2$, and maximum radius $r_\mr{max}=1$.
For the mass fraction of hot gas in the halo $f_\mr{gas}$ (sometimes called $f_\mr{hot}$), we use the mass dependent values from Table 3 of \citet{Khrykin2024} as a step function.
None of the galaxies contribute to the DM of the FRB, or rather likely none of the galaxy haloes are intersected by the FRB sightline.

However, the galaxy group LGG~93 has an impact parameter of \SI{0.54}{\Mpc}.
\citet{Marino2010} report a harmonic radius of \SI{0.46\pm0.04}{\Mpc}, a stellar mass of \SI{\sim4e9}{\Msol} and virial and projected total mass of \SI{1.96 \pm 0.44e13}{\Msol} and \SI{7.6\pm 5.2 e13}{\Msol}, respectively.
The virial and projected total masses yield a DM of $9^{+8}_{-4}\,\si{\dmu}$ and $31^{+29}_{-14}\,\si{\dmu}$, respectively, where the uncertainties are calculated from the uncertainties of $f_\mr{gas}$ given by \citet{Khrykin2024a}.
For our DM budget, we adopt the mean of $\DMX=21^{+15}_{-8}\,\si{\dmu}$.

\subsubsection{DM budget}

To estimate the host DM, $\DMh$, we subtract all other contributions from the measured \DM.
The Milky Way contribution \DMMW is obtained from the models NE2025 \citep{Cordes2002,Ocker2026} and YMW16 \citep{Yao2017}.
The contribution from the diffuse inter-galactic medium \DMd is estimated from the Macquart relation via the \tool{frb} package\footnote{\url{https://github.com/FRBs/FRB}}.
Assuming a constant MW halo DM $\DMhalo=\SI{30}{\dmu}$ \citep{Cook2023}, we calculate
\begin{equation}
    \DMh = \DM-\DMMW-\DMhalo-\DMd-\DMX\,.\label{eq:DMh}
\end{equation}
The resulting $\DM_\mr{host,rest}=224^{+17}_{-30}\,\si{\dmu}$ is relatively high compared to the median of the population of $105^{+61}_{-51}\,\si{\dmu}$ \citep{Hoffmann2025} but within the expected 68 per cent interval of a log normal population.
However, it is very high for the low inclination of its host and the spectral line ratios that resemble a MW like inter stellar medium.

\subsubsection{PRS Search}

Motivated by the high scattering and proximity of \FRBt, we searched for a persistent radio source at its location. Observations were made using the Australia Telescope Compact Array (ATCA), with the Compact Array Broadband Backend \citep[CABB;][]{Wilson2011}. We observed for \SI{5.5}{h} in the 4-cm band, covering \SI{5}{\GHz} to \SI{6}{\GHz} and \SI{8.5}{\GHz} to \SI{9.5}{\GHz}. Image reduction was performed using \tool{Miriad} \citep{Sault1995}. In both bands, we reached a root-mean-square of \SI{7}{\micro\Jy\per beam}, giving a 3-sigma upper limit of \SI{21}{\micro\Jy} corresponding to a specific luminosity of $L_\nu<\SI{1.2e27}{\erg\per\Hz\per\s}$, and $\nu L_\nu<\SI{6.7e36}{\erg\per\s}$ and $\nu L_\nu<\SI{1e37}{\erg\per\s}$ at 5.5 and \SI{9}{\GHz}, respectively.
This limit is about 70 times lower than the PRS of FRB\,121102 \citep{Marcote2017,Bhardwaj2025} and 120 times lower than that of FRB\,190520B \citep{Niu2022}.

\subsection{\FRBo}

For \FRBo, the DECam legacy survey showed a faint blob with r-band magnitude of 22.36.
Because this did not allow for a high probability host identification using PATH \citep{Aggarwal2021a}, we conducted dedicated follow-up imaging observations of the \FRBo location with the FORS2 instrument on the VLT (\texttt{Program ID 108.21ZF}, \texttt{PI Shannon}). The obtained special R band image was reduced following the pipeline described in \citet{Marnoch2023}. We conducted spectroscopic observations with the X-Shooter spectrograph on the VLT \citep{Vernet2011} under the same \texttt{Program ID 108.21ZF} (\texttt{PI Shannon}). A clear detection of H$\alpha$ plus faint traces of [O\,\textsc{ii}]\,$\lambda3727$ and [O\,\textsc{iii}]\,$\lambda5007$ yield a redshift of $z=0.3247\pm0.0003$.

The host galaxy is shown in the left panel of \cref{fig:HGs}, and shows no discernible structure at the given resolution.

We searched for intervening galaxies as described in \cref{sec:frbtinterv} but for galaxy redshifts $z<0.33$.
There is only one intervening galaxy with impact parameter $<\SI{0.3}{\Mpc}$, GLADE+ 5951486, with an impact parameter of \SI{58\pm 9}{\kpc} and $z=0.20\pm0.04$. Scaling the stellar mass of \SI{2\pm1e10}{\Msol} in GLADE+ to the total mass via a scaling relation \citep{Moster2010, Girelli2020}, yields \SI{5\pm2e11}{\Msol} and following \cref{sec:frbtinterv}, a DM contribution of $\DMX=15^{+3}_{-4}\,\si{\dmu}$.

Using \cref{eq:DMh} and following the surrounding description, we obtain a host DM of $\DM_\mr{host,rest}=106_{-106}^{+55}\,\si{\dmu}$.
This value is centred around the median of the population but also consistent with a much lower value.

\section{Discussion}
\label{sec:discussion}

\begin{figure}
    \centering
    \includegraphics[width=\linewidth]{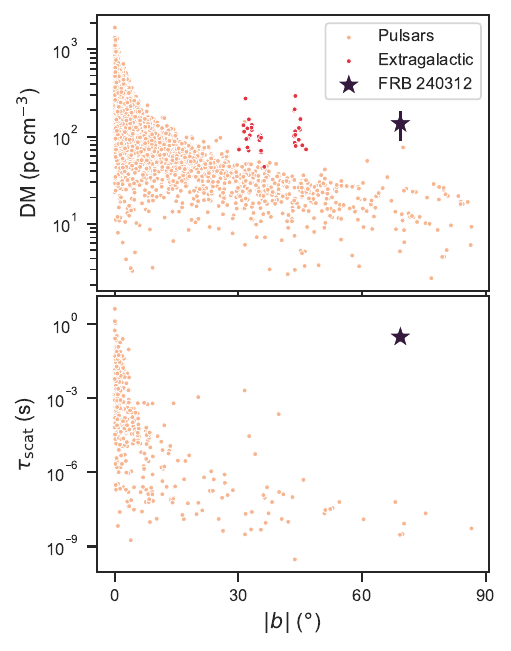}
    \caption{The \DMh, scatter broadening and host galaxy inclination of \FRBt compared to the population of pulsars in the MW and their Galactic latitude.}
    \label{fig:DMb}
\end{figure}

\subsection{\FRBt}
\FRBt shows a very high scattering time and DM given its location within its host galaxy.
The FRB's scintillation, presumably caused by a screen in the MW, constrains the scattering screen to be in the host galaxy.
Likewise, a large amount of the DM, $\DMh=225^{+17}_{-30}\,\si{\dmu}$, originates in the host.
The scattering and DM values are at odds with the low host galaxy inclination, its MW-like ISM properties and the FRB's offset from the centre.
It is therefore likely that the scattering and DM have their origin in the same ionized, inhomogeneous structure.

\Cref{fig:DMb} shows the comparison of \FRBt{}'s DM and \taus at its host inclination to the pulsar population at different Galactic latitudes, taken from the ATNF Pulsar Catalogue \citep{Manchester2005}.
The only sources in the MW where we observe a comparable scattering to \FRBt are sources viewed through several kpc of the inner disc or located in the Galactic centre.
Therefore, the only single region in the MW that has a measured scattering time comparable to \FRBt is the Galactic centre.

The lack of MW analogs, the clear offset from its host galaxy centre, and the constraint from \cref{eq:dist_prod} on the screen distance of $\Dhs\lesssim8\pm\SI{2}{\pc}\,(D_\mathrm{MW}/\si{\kpc})^{-1}$ strongly suggest that the screen must be in the circumsource medium.

\subsection{Screen models}
Observations in the past decades and the development of scintillometry have changed the interpretations of scattering in pulsars \citep{Stinebring2001}.
Established models explained pulsar scintillation by diffraction in the turbulent ISM along the line of sight, with occasional refraction into two or three images \citep{Lee1975,Rickett1977}.
Observations of scintillation arcs in the Fourier transform of their dynamic spectra showed that scattering must instead originate from hundreds of refractive images such that the refractive component dominates the observed scintillation \citep{Stinebring2001}. 

New interpretations of pulsar scintillation invoke refractive effects at larger scales \citep{Goldreich2006,Pen2014}.
New theoretical techniques allow exact solutions of involved integrals, which give exact solutions for some simple lenses \citep{Feldbrugge2019,Jow2023}.
Nevertheless, it is challenging to draw
conclusions on e.g. the ISM properties.

It is difficult to distinguish between refractive and diffractive scattering for a single FRB having only the scattering timescale measured.
While \citet{Jow2023} argue that strong scintillation is always in the refractive regime, it is less clear for scattering.
The Fresnel scale of a screen is typically used to decide if diffractive or refractive theory is applicable \citep{Rickett1990}.
If the length scale $a$ of plasma fluctuations or plasma lenses is $a\gg r_\mr{F}$, the deflection is refractive, while if $a\ll r_\mr{F}$, it is diffractive.

The Fresnel scale of a screen in the host galaxy can be derived from the general cosmological case \citep{Macquart2004} as
\begin{alignat}{1}
    r_\mr{F}&=\left(\frac{\Dh\Dhs}{\Ds}\frac{\lambda_\mr{obs}}{2\pi(1+z)}\right)^{1/2}\\
    &\approx\left({\Dhs}\frac{\lambda_\mr{obs}}{2\pi(1+z)}\right)^{1/2}\\
    &=\SI{3.7e9}{\cm}\left(\frac{\Dhs}{\SI{1}{\pc}}\right)^{1/2}\left(\frac{\nu(1+z)}{\SI{1}{\GHz}}\right)^{-1/2}\,,
\end{alignat}
where $\Dh$ and $\Ds$ are the angular diameter distance from the observer to the scattering screen in the host, and to the FRB source, respectively, and $\lambda_\mr{obs}$ is the observing wavelength.
In the case of \FRBt, the scale is $r_\mr{F}=\SI{4.2e9}{\cm}\left({\Dhs}/{\SI{1}{\pc}}\right)^{1/2}$.
This small length scale makes it challenging to generate diffractive scintillation, but it is not impossible, as we will discuss below.
As we will see, the characteristic scales of refraction and diffraction that we infer (\cref{eq:x_lim} and \cref{eq:rdiff}) are on their respective side of the Fresnel scale.
Both theories would therefore be self-consistent.

We note \citet{Jow2023} show that the relevant quantity is $r_\mr{F}/\sqrt{\kappa}$ rather than $r_\mr{F}$, where $\kappa$ is the convergence or lens strength, but it is not clear how to apply this to a screen that is not a single lens.

Since it is unclear if the scattering process is refractive or diffractive, we try to be agnostic and apply the different available theories, including refractive scattering and two models of diffractive scattering.

Without assuming anything about the nature of the scattering screen, we can derive a few quantities that give us an idea of the scales involved.
The angular size of the FRB image when $\Ds\approx\Dh$ is \citep{Pradeep2025}
\begin{alignat}{1}
    \thh&\approx\left(\frac{2c}{1+z}\frac{\Dhs}{\Ds\Dh}\taus\right)^{1/2}\nonumber\\
    &=\SI{1e-7}{arcsec}\left(\frac{\Dhs}{\SI{1}{\pc}}\right)^{1/2}\,,\label{eq:thetahost}
\end{alignat}
where $c$ is the speed of light.
The transverse radius of the illuminated screen is then
\begin{equation}
    X_\mr{host}=\Dh\thh=\SI{23}{\astronomicalunit}\left(\frac{\Dhs}{\SI{1}{\pc}}\right)^{1/2}\,.\label{eq:Xhost}
\end{equation}
We can calculate the angle of deflection that is caused by the screen as
\begin{alignat}{1}
    \alpha_\mr{d}&=\frac{X_\mr{host}}{\Dhs}+\thh=\left(\frac{\Dh}{\Dhs}+1\right)\thh\\
    &\approx\frac{\Ds}{\Dhs}\thh\,
\end{alignat}
where the last approximation is only valid for $\Dhs\ll\Dh$, since generally $\Ds\neq\Dh+\Dhs$ in a non-Euclidean universe.
Using \cref{eq:thetahost}, we obtain
\begin{equation}
    \alpha_\mr{d}\approx\left(\frac{2c\taus}{(1+z)\Dhs}\right)^{1/2}=\SI{23}{arcsec}\left(\frac{\Dhs}{\SI{1}{\pc}}\right)^{-1/2}\,.\label{eq:alpha}
\end{equation}

\subsubsection{Refractive}
If the scattering process is refractive in nature, the refraction is caused by inhomogeneities in the electron density.
We can therefore relate it to the gradient in DM perpendicular to the line of sight
using the stationary phase condition \citep{Gwinn1998,Walker2004}.
Because distances are cosmology dependent while angles are not, we express the stationary phase condition in angular space:
\begin{equation}
    \frac{\partial\Phi}{\partial\theta} = 0
\end{equation}
The phase delay $\Phi$ of each path is given by the dispersive phase delay quantified by its \DM and the geometric delay quantified by $\taus$:
\begin{equation}
    \Phi = -2\pi a\times\frac{\DM}{\nu} + 2\pi\nu \taus
\end{equation}
where $a\approx\SI{4.15}{\ms\GHz\squared\per\pc\cm\cubed}$ \citep{Lorimer2004}. The relation between scattering angle and scattering delay for a screen in the host galaxy is given in \cref{eq:thetahost}. Since the scattering time corresponds to the characteristic scattering angle, the characteristic angular dispersion measure gradient is determined by the stationary phase condition:
\begin{alignat}{1}
    \langle|\partial_\theta\DM|\rangle&=\frac{2\nu^2}{ a}\left(\frac{1+z}{2c}\frac{\Ds\Dh}{\Dhs}\taus\right)^{1/2}
\end{alignat}
Expressing the gradient in terms of the proper transverse length at the screen $x=\theta\Dh$, we get
\begin{alignat}{1}
    \langle|\partial_x\DM|\rangle&=\frac{\langle|\partial_\theta\DM|\rangle}{\Dh}\\
    &=\frac{2\nu^2}{ a}\left(\frac{1+z}{2c}\frac{\Ds}{\Dh\Dhs}\taus\right)^{1/2}\\
    &=9.3\,\si{\pc\per\cm\cubed\per\astronomicalunit}\left(\frac{\Dhs}{\SI{1}{\pc}}\right)^{-1/2}\,.\label{eq:DMgrad}
\end{alignat}
Possible dimensions range from one large structure of size $2X_\mr{host}$ to very small structures.
However, the largest extent with a gradient of $\Delta\DM=\partial_x\DM\cdot 2X_\mr{host}=\SI{422}{\dmu}$ yields a \DM that is higher than the observed \DM.

Still, we need to take into account that a large spread in DMs could be hidden under the scattering tail.
We can estimate an upper limit on \DM variations from the observed width of the FRB.
Given the delay due to \DM as $\Delta t=a\DM\nu^{-2}$ and considering that delay differences must be smaller than the observed width $w_t$, i.e., $\Delta t\lesssim w_t\approx\taus$, we can set an approximate upper limit of $\Delta\DM\lesssim\taus \nu^2/a=\SI{108}{\dmu}$.
Assuming giant wedges with a linear gradient and length $x$, this yields
\begin{equation}
    x\lesssim\frac{\Delta\DM}{\langle|\partial_x\DM|\rangle} =0.5\cdot X_\mr{host}\,.\label{eq:x_lim}
\end{equation}

\subsubsection{Diffractive}
If the FRB is scattered diffractively, there is a relationship between the deflection angle and the diffractive length scale $r_\mr{diff}=\lambda/2\pi\alpha_\mr{d}$, where $\lambda$ is the rest frame wavelength \citep[][]{Blandford1985,Rickett1990}.
With \cref{eq:alpha} and $\lambda=\lambda_\mr{obs}/(1+z)$, we obtain
\begin{alignat}{1}
    r_\mr{diff}&=\frac{1}{2\pi\nu}\sqrt{\frac{c}{2\taus(1+z)}\Dhs}\\
    &=\SI{5e4}{\cm}\left(\frac{\Dhs}{\SI{1}{\pc}}\right)^{1/2}\,. \label{eq:rdiff} 
\end{alignat}

The inhomogeneity in the medium is typically thought to come from turbulence.
The turbulence is modelled by a power-law in length scales.
It is thought to be injected at a large length scale $L_0$ and cascades down to small length scales until a minimum $l_0$.
Here, we will follow the formulation of \citet{Yang2022a} and assume a shell around the source with radius $R$ and shell thickness $\Delta R$.

In the model, $r_\mr{diff}$ depends on two case distinctions, whether $r_\mr{diff}$ is less or greater than the lowest scale of turbulence $l_0$, and whether the turbulence spectral index $\beta$ is less or greater 3; for Kolmogorov turbulence $\beta=11/3$, hence we will assume $\beta>3$.
In this case, we get
\begin{equation}
    r_\mr{diff}=\begin{cases}
    (A_1r_e^2\lambda^2(l_0/L_0)^{\beta-4}L_0^{-1}\delta n_{e,0}^2\Delta R)^{-\frac{1}{2}} & r_\mr{diff}<l_0\\
    (A_2r_e^2\lambda^2L_0^{3-\beta}\delta n_{e,0}^2\Delta R)^{\frac{1}{2-\beta}}& r_\mr{diff}>l_0
    \end{cases}\,,\label{eq:rdiff_med}
\end{equation}
where $r_e$ is the classical electron radius and $\delta n_{e,0}^2$ is the total mean squared density fluctuation,

\begin{alignat}{1}
A_1&=\Gamma(2-\beta/2)\pi^2\frac{\beta-3}{(2\pi)^{4-\beta}}\approx 10\,, \quad\text{and}\\
A_2&=\frac{\Gamma(2-\beta/2)}{\Gamma(\beta/2)}\frac{8\pi^2}{(\beta-2)2^{\beta-2}}\frac{\beta-3}{2(2\pi)^{4-\beta}}\approx 16\,,
\end{alignat}
with the gamma function $\Gamma$ and the numerical values for $\beta=11/3$.

\subsubsection{Densities}
\label{sec:densities}

\begin{figure}
    \centering
    \includegraphics[width=1\linewidth]{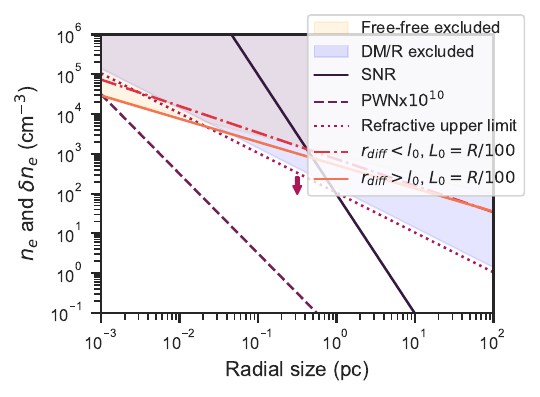}
    \caption{Allowed and required densities at different source-screen distances for different models, assuming a homogeneous density around the source. For the first four legend entries, $n_e$ is shown and for the last three $\delta n_e$.}
    \label{fig:density}
\end{figure}

We want to compare the density variations needed in the refractive and diffractive case to the densities in two simple models, a homogeneous sphere and a homogeneous shell.
We will discuss these specifically in the context of a SNR and a pulsar wind (PW) as was previously done by \citet{Kumar2024}.
We show the different density limits calculated in this section, in \cref{fig:density}.

The electron number density in a homogeneous, ionized sphere with radius $R$ can be modelled as
\begin{equation}
    n_e(R)=\frac{M_p}{m_p}\frac{1}{\frac{4}{3}\pi R^3}=\SI{9.7}{\per\cm\cubed}\frac{M_p}{\SI{1}{\Msol}}\left(\frac{R}{\SI{1}{\pc}}\right)^{-3}\,,\label{eq:ne_sphere}
\end{equation}
where $M_p$ is the mass of all protons in the sphere and $m_p$ is the proton mass.

As discussed by \citet{Kumar2024}, the most stringent upper limits come from free-free absorption and from the observed dispersion measure.
The free-free optical depth is given by \citep{Rybicki1986}
\begin{equation}
    \tau_\mr{ff}=\alpha_\mr{ff}\Delta R=A_\mr{ff}T^{-\frac{3}{2}} Z^2 n_e n_i \nu^{-2}\bar{g}_\mr{ff}\Delta R\,,\label{eq:tauff}
\end{equation}
where $\alpha_\mr{ff}$ denotes the absorption coefficient, $A_\mr{ff}=\SI{0.018}{\cm^5 \kelvin^{3/2} \per \s\squared}$ is a combination of physical constants, $T$ is the gas temperature, $Z$ is the ion charge number, $n_i$ is the ion density, and $\bar{g}_\mr{ff}$ is the frequency dependent gaunt factor which takes the value $\bar{g}_\mr{ff}\approx 7$ \citep{Chluba2019}.
Since the FRB was able to escape, we must have $\tau_\mr{ff}\lesssim 0.5$ and assuming $Z=1$ and $n_i=n_e$, \cref{eq:tauff} yields
\begin{equation}
    n_e\lesssim \sqrt{\frac{0.5 \,T^{\frac{3}{2}} \nu^{2}}{A_\mr{ff}\, \bar{g}_\mr{ff} \Delta R}}=\SI{900}{\per\cm\cubed}\left(\frac{T}{\SI{e4}{\K}}\right)^{\frac{3}{4}}\left(\frac{\Delta R}{\SI{1}{\pc}}\right)^{-\frac{1}{2}}\,.\label{eq:neff}
\end{equation}

Parts of the DM could originate in the ISM, therefore the DM can only serve as another upper limit on the density. Attributing at least \SI{40}{\dmu} to the host halo and host ISM each yields
\begin{equation}
    n_e\lesssim\frac{\DMs}{\Delta R}\approx \SI{140}{\per\cm\cubed}\frac{\SI{1}{\pc}}{\Delta R}\,.\label{eq:ne_DM}
\end{equation}

In the context of a homogeneous SNR ($\Delta R=R$), the remnant must be significantly expanded. Combining \cref{eq:ne_sphere} and \cref{eq:ne_DM}, we obtain
\begin{equation}
    R\gtrsim \sqrt{\frac{M_p}{\frac{4}{3}\pi m_p\DMs}}=\SI{0.26}{\pc}\left(\frac{M_p}{\SI{1}{\Msol}}\right)^{\frac{1}{2}}\,.
\end{equation}
The SNR age follows as
\begin{equation}
    t_\mr{age}\gtrsim\SI{26}{\yr}\left(\frac{M_p}{\SI{1}{\Msol}}\right)^{\frac{1}{2}}\left(\frac{v_\mr{ej}}{\SI{e4}{\km\per\s}}\right)^{-1}\,,
\end{equation}
where $v_\mr{ej}$ is the expansion velocity of the SNR.

We want to compare the obtained density limits with the density fluctuations that are required for the observed scattering.
Under the assumption that the scattering is refractive, we can only obtain an upper limit for the density variations. Assuming a linear gradient of the DM,  $\langle|\partial_x\DM|\rangle=\Delta n_e\Delta R/x$ and since we have an upper limit on $x$ from \cref{eq:x_lim}, we combine it with \cref{eq:Xhost} and \cref{eq:DMgrad} and get
\begin{equation}
   \Delta n_e=\frac{\langle|\partial_x\DM|\rangle\,x}{\Delta R} \lesssim \frac{\Delta\DM}{\Delta R} = \SI{108}{\per\cm\cubed}\frac{\SI{1}{\pc}}{\Delta R}\,.
\end{equation}

For the diffractive case, we solve \cref{eq:rdiff_med} for $\delta n_{e,0}$, which yields (for $\beta=11/3$)
\begin{equation}
    \delta n_{e,0}=\begin{cases}
    (r_e\lambda r_\mr{diff})^{-1}\left(\frac{l_0}{L_0}\right)^{0.17}\left(\frac{L_0}{A_1\Delta R}\right)^{\frac{1}{2}} & r_\mr{diff}<l_0\\
    (r_e\lambda)^{-1}r_\mr{diff}^{-0.83}L_0^{0.33}(A_2\Delta R)^{-\frac{1}{2}}& r_\mr{diff}>l_0
    \end{cases}\,.
\end{equation}
In \cref{fig:density}, we fix $\Delta R=R$, $l_0=\SI{e5}{\cm}$ in the first case, and $L_0=R/100$.

\Cref{fig:density} shows that the upper limit from \DM almost coincides with the observational upper limit in the refractive case while free-free absorption only limits it lightly at small radii.
We can therefore put almost no limit on the physical properties of the medium that could cause such refraction.
In other words, a wide range of density-length scale combinations could produce the amount of scattering we observe.
Still, a physical model needs to explain the clustering of matter on scales of $\lesssim 0.5\,X_\mr{host}$.

On the other hand, it seems challenging to obtain the density fluctuations required for diffractive scintillation.
The only sweet spot is for $r_\mr{diff}\sim \SI{2e-2}{\pc}$, though excluded in the case of an SNR.
The only other possibility would be that $\delta n_{e,0}\gg n_e$.
Interestingly, this illustrates that also generally, diffraction requires a significant amount of DM.

\subsubsection{Cloudlet model}
In the diffractive case, we can further apply the cloudlet model that was developed based on pulsar observations \citep[see][and references therein]{Cordes1991,Cordes2022}.
If the screen is close to the source, the geometric factor $G$ in their model becomes $G\approx 1$.
Applying the model of \citet{Cordes2022}, yields a fluctuation parameter of $\tilde{F}=\SI{154}{(\pc\squared\km)^{-1/3}}$, well above the highest value considered reasonable by \citet{Cordes2022} of $\tilde{F}=\SI{10}{(\pc\squared\km)^{-1/3}}$.
Put in a different way, the \DM of the scattering medium would have to be above \SI{800}{\dmu} to give reasonable parameter values.

\subsubsection{Characteristic length scales}

It is unclear which physical scales determine the minimum length scale of turbulence $l_0$.
\citet{Ocker2025} discuss five different physical length scales.
Under the assumption that $r_\mr{diff}$ cannot be much smaller than $l_0$, we can calculate the physical conditions required to get physical length scales of the same order.

The ion inertial scale $\lambda_i=\SI{230}{\km}\,(n_e/\si{\per\cm\cubed})^{-1/2}$ would imply a density of $n_e\sim\SI{2e5}{\per\cm\cubed}$.
This is much larger than the density allowed by free-free absorption (\cref{eq:neff}).

The gyroradius of ions might instead set $l_0$.
This would require a magnetic field of \citep{Draine2011}
\begin{alignat}{1}
    B&\sim\frac{c}{e\,r_\mr{diff}}\sqrt{m_\mr{p}k_\mr{B}T}\\
    &=\SI{2}{\milli\gauss}\left(\frac{T}{\SI{e4}{\K}}\right)^{1/2}\left(\frac{\Dhs}{\SI{1}{\pc}}\right)^{-1/2}\label{eq:Blim}\,.
\end{alignat}
and for electrons $B\sim\SI{42}{\micro\gauss}\left(\frac{T}{\SI{e4}{\K}}\right)^{1/2}\left(\frac{\Dhs}{\SI{1}{\pc}}\right)^{-1/2}$.
In the model of \citet{Yang2022}, the luminosity of a PRS in terms of B is given in their eq.~(43). Combining it with \cref{eq:Blim}, we obtain
\begin{alignat}{1}
    L_\nu\gtrsim  \SI{1.5e28}{\erg\per\s\per\Hz}\left(\frac{\nu}{\SI{5.5}{GHz}}\right)^{-1/2}\\
    \times\left(\frac{T}{\SI{e4}{\K}}\right)^{7/4}\left(\frac{\Dhs}{\SI{1}{\pc}}\right)^{5/4}\,.\nonumber
\end{alignat}
This would only be consistent with our upper limit of $L_\nu<\SI{1.2e27}{\erg\per\s\per\Hz}$ if $\Dhs\lesssim \SI{0.13}{\pc}$.

Other length scales that could influence turbulence are even smaller.
So they are allowed by the data, but we cannot draw any useful constraints in those cases.

\subsection{Possible scattering origins}

Given the constraints on the screen, we discuss several possible physical models.
These are largely motivated by the models for the PRSs of FRBs.
In summary, the observables that need to be explained are:
\begin{itemize}
    \item the large scattering of $\taus=\SI{0.7}{\s}$
    \item the close distance of $\Dhs\lesssim\SI{8}{\pc}\,(D_\mathrm{MW}/\si{\kpc})^{-1}$ constrained by scintillation
    \item the location in a spiral arm \citep{Gordon2025}
    \item the presence of a bright star-forming region at the FRB location
    \item the absence of signatures of any other objects e.g. a supermassive black hole
    \item the upper limit on a potential PRS luminosity.
\end{itemize}
The process of scattering is likely refractive with a DM gradient of $9\,\si{\pc\per\cm\cubed\per\astronomicalunit}\left(\frac{\Dhs}{\SI{1}{\pc}}\right)^{-1/2}$ (see \cref{eq:DMgrad}).

\subsubsection{Pulsar wind nebula}
A young neutron star (NS) is a common FRB model \citep[see][for reviews]{Platts2019,Petroff2019}.
If embedded in an SNR, the SNR would have the required density and the NS wind could cause the required inhomogeneous structures.
Pulsars and magnetars have strong relativistic winds of electrons and positrons.
A few decades after the SN, the wind that drives a shock into the SNR can form a PWN \citep[see][for a review]{Reynolds2017}.
At the accelerating shock front of the expanding PWN, a thin shell of swept up gas builds up.
Rayleigh-Taylor instabilities in the swept up shell cause the formation of inward pointing `fingers' and less prominent filaments between these fingers \citep[e.g.][and references therein]{Hester1996, Hester2008}.
A layered structure allows these fingers to persist in the PW and the inner gas to cool and become denser.
They are expected to develop \SIrange{100}{200}{\yr} after the supernova \citep{Jun1998}.
Filaments are believed to reach densities of \SIrange{1000}{2000}{\per\cm\cubed} \citep{Sankrit1998}.
Simulations reach a resolution of \SI{330}{\astronomicalunit}, at least an order of magnitude higher than the structures that could produce our scattering, but the dependence on resolution suggests a continuation to smaller scales \citep{Porth2014}.

The Crab pulsar (PSR~B0531+21) is the prime example of a young pulsar with a PWN.
Crab observations show scattering and scintillation originating in the PWN with strong temporal variations \citep{Rankin1970,Vandenberg1976,Main2021}.
However, the scattering timescale of the Crab pulsar only reaches \SI{0.08}{\ms} at \SI{1}{\GHz} \citep[scaled from \SI{0.6}{\ms} at \SI{610}{\MHz};][]{McKee2018}, a factor 3500 less than we observed for \FRBt.
So how does scattering scale with the PWN properties?

It is not clear how the PWN's ability to scatter evolves with age.
Assuming the PWN accelerates with $v\propto t^{6/5}$ \citep{Olmi2023} the average DM imposed by the shell slowly decreases with $\propto t^{-4/5}$, and scattering is further diminished by the geometric $R^{1/2}$.
This means that just scaling the Crab to a younger age is not sufficient to make this model work.
Rather, the sightline has to be in a much denser direction than the one of the Crab pulsar, for example right through a filament.
Another possibility is that there is a time frame in the evolution where the neutral molecular structures within the filaments of the present Crab PWN were already dense but still ionized.

This model is likely not working for strong magnetar winds.
If the wind is too strong, it would push through the shell as proposed for the case of super luminous supernovae.
Without the shell, filamentary structure could not form in the same way as in a normal PW.

Despite the gap between the observed scattering in our FRB and in PWNe, it is one of few plausible options.
In this case, the observed star-forming region at the FRB position is not related to the scattering.
However, its presence is a prerequisite for a young PWN.

\subsubsection{Massive black holes}
Supermassive or intermediate mass black holes can accrete substantial amounts of material that could result in high DM and $\taus$ in a nearby or passing FRB.
The Galactic centre is estimated to cause up to several seconds of scatter broadening on pulsars within it \citep[see e.g. the discussion in][]{Ocker2026}.
Wandering black holes offset from their galaxy centres are expected from simulations \citep{Ricarte2021} and have been observed \citep[see e.g.][]{Mezcua2024}.
A massive black hole has been invoked as the origin of persistent radio sources in several scenarios \citep{Romero2016,Katz2017,Zhang2017,Zhang2018b}.

Our follow-up observations put strong constraints on this model.
Our integral field spectroscopic analysis shows no evidence of a hard ionizing radiation spectrum, excluding the presence of a supermassive black hole at the FRB position and even in the galactic centre.
The nondetection of a PRS with ATCA further reinforces this conclusion \citep{Morabito2022}.
Nevertheless, a smaller perhaps intermediate mass black hole cannot be excluded at the given distance.

\subsubsection{High Mass Companion star}

In a neutron star--massive companion star binary system, a strongly magnetised, slowly spinning neutron star accretes from a massive companion star (typically OB supergiants or Be stars with strong stellar winds). Coherent radiation from the neutron star may be scattered and depolarised when passing through the Be star decretion disc \citep{Wang2022a, Rajwade2023}. The dense stellar material modulates when bursts are seen and introduces strong propagation effects to the burst morphology.

For such scenarios, a strong rotation measure variation, repeating activity would be strong additional evidence. Persistent radio counterparts may also be expected from synchrotron emission generated by shocks in the companion star stellar wind. For a typical $\gamma$-ray binary, the expected luminosity of \SI{3e20}{\erg \per\s \per\Hz} is orders below the ATCA limit, agreeing with the PRS non-detection.

For \FRBo and \FRBt, the scattering time is much larger than for FRB\,201124A \citep{Wang2022a}, thus requiring 3-15 times higher $\delta n_e$ and hence $\delta n_e > 10^7 \mathrm{cm^{-3}}$. Such densities would not support an optical depth sufficient for emission to be visible. 

\subsubsection{Hypernebula}
\citet{Sridhar2022} proposed a model where the FRB and persistent radio emission stem from a neutron star or black hole in a binary with a massive post-main sequence star.
When the companion star expands, it fills its Roche lobe and the overflowing matter is accreted on to the compact object.
Super Eddington accretion at the start of this process results in winds from the accretion disc that shock the surrounding medium producing a `hypernebula'.
The model includes \num{\sim10} parameters, and makes predictions of only two of our measured parameters: the DM and PRS luminosity.
We can therefore not hope to fit the model, but we can read the order of magnitude of the age from the example model plots \citep[fig.~6 in][]{Sridhar2022}.
Considering our source DM is of order \SI{140}{\dmu} (see \cref{sec:densities}) and we obtained an upper limit on the PRS luminosity of $\nu L_\nu<\SI{1e37}{\erg\per\s}$, the age of the hypernebula would be \SI{\sim1}{\yr}.
In a younger system the DM would be higher, while in an older system the PRS would be visible.
Apart from the age the most important parameters are the velocities of the jet and wind.
A lower wind velocity might delay the luminosity peak by tens of years allowing for a much older system, while a lower jet velocity could allow the peak to have already passed.
In both cases, the nebula could be tens of years old.

\subsubsection{\HII regions}
\HII regions ionized by massive stars are common in the MW and star-forming galaxies.
Intersections of pulsar sightlines with \HII regions can explain some high pulsar DMs \citep{Ocker2024}.
But individual regions contribute $\taus[,\SI{1}{\GHz}]\ll\SI{100}{\ms}$.
\HII regions can therefore be rejected relatively clearly.

Cygnus-X is one of the largest star-forming regions in the MW.
\citet{Patil2026} recently reported a detection gap of FRBs in the CHIME catalogue in the Cygnus-X sky region.
They found the foremost reason to be scatter broadening and put a lower limit of $\taus[,\SI{1}{\GHz}]\geq\SI{5.59}{\ms}$.
Scattering of other sources behind the region has been previously estimated \citep[see the discussion by][]{Patil2026}.
The highest measured scattering was $\taus[,\SI{1}{\GHz}]=\SI{127}{\ms}$ \citep{Jing2025}.
While the source distance of the pulsar is unknown, galactic sources have a large leverage compared to an extragalactic source, which would be less scattered by the same region.
Nevertheless, the scattering time is only a factor of a few below the one of \FRBt.

Despite the scattering by Cygnus-X being below our measured scattering, there could be sightlines with comparable scattering.
Similar to the PWN case, such a model would require a high degree of finetuning of the relative positions to get a high-scattering sightline but also of the short distance from the FRB source to the scattering medium.
This finetuning implies a large population of FRBs at comparable sites that are less scattered.
This is so far disfavoured by the typical FRB positions in hosts, the majority of which show no clear preference for star formation regions \citep{Gordon2025}.
Another counterargument is that Cygnus X is one of the brightest Galactic \HII regions, while several \HII regions in the FRB host are brighter in H$\alpha$ than the one at the FRB position.
    
\subsection{\FRBo}

We can say much less about the scattering origin of \FRBo.
The host galaxy lies at a much greater distance, and the FRB cannot be pinpointed to a specific location within it.
It might, for instance, be situated near the centre, which could account for its strong scattering.

We consider the constraints under the different scattering models.
Under the assumption of refractive scattering, the FRB is less constraining than \FRBt because the constraints depend only on \taus.
The same is true for the diffractive case.
However, the cloudlet model \citep{Cordes1991,Cordes2022} is DM dependent.
Applying it to this FRB, we find the fluctuation parameter $\tilde{F}=180^{+50}_{-90}\,\si{(\pc\squared\km)^{-1/3}}$.
The value is higher than in \FRBt because of the lower \DMh but the larger uncertainties make it less restrictive.

If the screen is instead in the halo of the intervening galaxy, the geometric factor becomes $G\sim 5000$ and the fluctuation parameter $\tilde{F}=\SI{1.3}{(\pc\squared\km)^{-1/3}}\,\left(\frac{L}{\SI{100}{\kpc}}\right)$, where $L$ is the screen thickness.

\citet{Cordes2022} consider $\tilde{F}$ values to be possible between 0.01 and $\SI{10}{(\pc\squared\km)^{-1/3}}$.
The case where the screen is in the intervening galaxy with a thickness of 1 to \SI{1000}{\kpc} would therefore give reasonable values, while the screen in the host galaxy cannot.

\subsection{Should surveys be optimised to find long duration FRBs?}
\begin{figure}
    \centering
    \includegraphics[width=\linewidth]{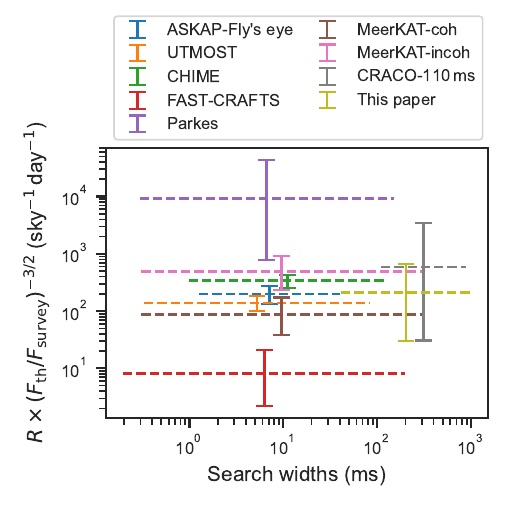}
    \caption{Rates observed by different telescopes and scaled to our fluence threshold assuming the power-law index in a Euclidean space. Dashed lines show the extent of the searched boxcar widths. The data are taken from \citet{James2019a,Gupta2022,CHIME2021,Niu2021}, \citet{Wang2025} \citep[re-estimated from][]{Crawford2022}, \citet{Jankowski2023} (twice), \citet{Wang2025}, respectively.}
    \label{fig:rate}
\end{figure}
Fig.~\ref{fig:rate} shows the measured rate compared to other surveys, where we scaled everything to a reference fluence of $F_\mr{ref}=\SI{9}{\Jy\ms}$.
Our measured rate agrees well with the observed rate during the CRACO 110-\si{\ms}-resolution pilot survey \citep{Wang2025} that probed the same range of FRB widths.
Furthermore, the rates of both agree or could even be higher -- per logarithmic width window -- than the rate at lower widths.

The two reported FRBs give additional evidence against a falling rate towards high width and high scattering timescales,
in favour of the statistical analysis of \citet{James2026}.

Because the results suggest that the rate is log uniform, it is in principle worth searching in this part of the parameter space.
Whether it is the most computationally efficient, assuming one seeks to maximise the detected number of FRBs, depends on the survey parameters.
Assuming each boxcar is probing a similar volume of parameter space (which seems reasonable after the first boxcar) with a rate $R_i=R_0(F_\mr{th}/R_0)^{-3/2}$ and if widths are powers of 2 -- like in the popular Heimdall search -- we get (applying $F_\mr{th}\propto\sqrt{w}$)
\begin{eqnarray}
    R_\mr{tot}=&&\sum_{i=0}^{N_\mr{bc}} R_0\left(\frac{w_i}{w_0}\right)^{-\frac{3}{4}}=R_0\sum_{i=0}^{N_\mr{bc}}2^{-i\frac{3}{4}}\nonumber\\
    =&&R_0\frac{2-2^{\frac{1}{4}(1-3N_\mr{bc})}}{2-\sqrt[4]{2}}\,,
\end{eqnarray}
where $N_\mr{bc}$ is the number of searched boxcars.
This function flattens rather quickly because the fluence threshold increases as $F_\mr{th}(i)\propto2^{i/2}$.
Therefore, long duration FRBs are not the most efficient to observe for most surveys that aim to use FRBs as tools but are very interesting for studying the origins of FRBs. 

\subsection{Can scattering be used as a \DMh estimate?}
\begin{figure}
    \centering
    \includegraphics[width=\linewidth]{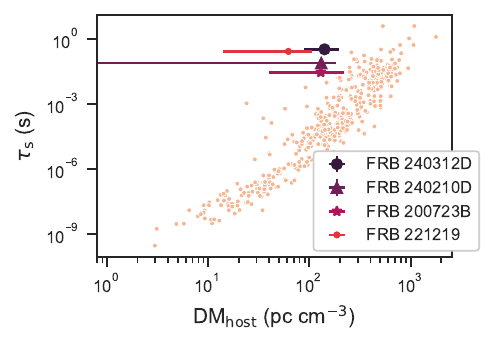}
    \caption{The FRBs compared to the known population of FRBs and pulsars. Data for \FRBfaber is taken from \citet{Faber2024} and \FRBshin from \citet{Shin2025}.}
    \label{fig:DMtau}
\end{figure}

It has been hoped since early in the FRB history that scattering in FRBs is related to \DMh \citep{Cordes2016}.
It is motivated by the empirical $\taus$-\DM relation seen in galactic pulsars \citep{Bhattacharya1992}, but it is unclear if FRBs should show a similar relation due to the possibly different ISM properties, the role of the circumburst medium, and the uncertain presence of extragalactic screens.
If found to be true, such a relationship would allow an estimate of \DMh, which is the biggest uncertainty in \DMd.

\FRBt shows high scattering and \DMh but the latter is not extraordinarily high compared to the population.
As shown in \cref{fig:DMtau}, this FRB is part of a growing population of highly scattered FRBs that are in disagreement with the relationship in pulsars.
It is currently unclear whether the relation exists at all or is just different from the one in the MW.
\FRBt is the first highly scattered FRB that has a precise enough \DMh to say that the DM is clearly above the average.
It therefore indicates that some relationship might exist but also shows that such a relationship would have a large spread.

Like previous FRBs, \FRBo shows high scattering and a comparably low DM.
However, in this case the evidence is less clear as the scattering origin is not known, and it is not completely secure if the FRB is actually scattered or just intrinsically wide.
To further address the question, we need to grow our population of nearby FRBs with well constrained \DMh to establish or discard a $\taus$-\DMh relationship \citep{Qiu2020,Scott2025, MasRibas2026}.

Similar conclusions can be drawn for the use of scattered FRBs as a probe of extragalactic ISMs \citep{Xu2016}.
Whether the host ISM can be studied depends on how often the source environment dominates the scattering, as in \FRBt.
Again, a population of nearby FRBs with scattering could give more insight.

\subsection{Can scattering be used to study intervening haloes?}

Observations and simulations show that the CGM consists of a multiphase gas \citep[see][for a review]{FaucherGiguere2023}.
Clouds in the CGM might act as scattering screens to traversing FRBs \citep{Prochaska2019a,Jow2024,MasRibas2025}.
This work illustrates that it could be difficult to associate scattering with an intervening screen.
For example, the argument of low DM used by \citet{Faber2024} is no longer a convincing indication for the association.
The reason is that the \DMh of \FRBt would also be consistent with the population if we had no inclination measurement for its host.
The best chance for an association might be from partially resolved sources that could be identified via the modulation index or low scattering power-law indices \citep[][\Bhatulap]{Pradeep2025}.
Additionally, upper limits \citep[see e.g.][]{Prochaska2019a,Ocker2021} are still possible.
The contribution of scattering by haloes can also be constrained statistically e.g. by getting the fraction of host scatterings from nearby FRBs and comparing them with correlations with sightlines of far-away FRBs with interveners.

\subsection{Implications for other highly scattered FRBs}

On a population level, our measured rate of highly scattered FRBs is about a quarter of what is observed at scattering times below \SI{10}{\ms} \citep{CHIME2021}.
This far exceeds the chance to have a highly scattered event in the MW at a random sightline.
Our measured rate is therefore indicative that high scattering in other FRBs \citep{Faber2024, Shin2025} is also not from coincident sightlines but is causally connected to the source of FRBs.

\section{Conclusions}
\label{sec:conclusion}

The two reported, highly scattered FRBs that we found with ASKAP-CRACO, have been challenging to explain.
In particular, \FRBt has an enigmatic origin and wide implications for the interpretation of highly scattered FRBs.
We draw the following conclusions:
\begin{itemize}
    \item \FRBt shows scintillation consistent with a screen in the MW and inconsistent with the scattering screen, implying a two-screen system.
    The resulting upper limit on the host screen distance of $\SI{8}{\pc}\,(D_\mathrm{MW}/\si{\kpc})^{-1}$ and the offset location of the FRB from its host galaxy centre strongly suggest that the scattering originates in the circumburst medium.
    \item Three points argue in favour of refraction over diffraction as a cause of the scattering:
    (i) the extent of the screen; (ii) the very low required diffractive scale; and (iii) the implied density variations close to densities where the emission would be free-free absorbed.
    \item Through integral field spectroscopy, we identify the knot at centre at the FRB localisation as a star-forming region. A supermassive black hole can be excluded, while other objects like an SNR or a PWN stay below our resolution and cannot be excluded.
    \item We find the origin of scattering with the most observational support to be the filamentary structure of a few hundred years old PWN, similar to what is seen in the Crab.
    The main difference from the Crab that causes the much larger scattering must mostly be a more inhomogeneous sightline e.g. through a filament, rather than age or mass in the SNR. Strong magnetar winds as in super luminous SNe where the wind breaks through the shell are less plausible in this scenario.
    \item Other models that have been proposed for PRSs of FRBs are possible as well within the observational constraints.
    A hypernebulae does not seem implausible either, although scattering has not yet been a focus of the models.  
    \item The finding that the high scattering of \FRBt can be traced back to the circumburst medium in this one FRB has important implications for the interpretation of other highly scattered FRBs. What used to be a very hypothetical possibility is now the most likely scenario for the scattering in e.g. \FRBo, \FRBshin \citep{Shin2025}, \FRBfaber \citep{Faber2024}.
    \item Large scattering from the circumburst medium poses problems for methods using scattering to study the host or MW ISM, or the CGM of intervening haloes. Furthermore, it questions the usability of scattering as an estimate for \DMh, the doubts are enforced by the relatively normal estimated \DMh seen in highly scattered FRBs.
    \item The rate calculated from the two reported FRBs is consistent with the rate seen at lower widths. This indicates the presence of a large population of highly scattered FRBs.
\end{itemize}

\section*{Acknowledgements}

JJ-S would like to thank Daniel Reardon, Ashley Stock, and Artem Tuntsov for helpful discussions about the origins of scattering and scintillation, Paz Beniamini for helpful discussions, in particular about the role of free-free absorption, Deanne Fisher for help with the integral field spectroscopy.

JJ-S, ATD, RMS, and YW acknowledge support through Australian Research Council Discovery Project DP220102305.

MG is supported through UK STFC Grant ST/Y001117/1. MG acknowledges support from the Inter-University Institute for Data Intensive Astronomy (IDIA). IDIA is a partnership of the University of Cape Town, the University of Pretoria and the University of the Western Cape. For the purpose of open access, the author has applied a Creative Commons Attribution (CC BY) licence to any Author Accepted Manuscript version arising from this submission.

A.C.G. and the Fong Group at Northwestern acknowledge support by the National Science Foundation under grant Nos. AST-1909358, AST-2206494, AST-2308182 and CAREER grant No. AST-2047919. A.C.G. acknowledges support from NSF grants AST-1911140, AST-1910471 and AST-2206490 as a member of the Fast and Fortunate for FRB Follow-up team.

NT acknowledges support from FONDECYT grant 1252229.

This paper benefitted from discussions at FTSky: A program in the field of Fast Radio Transients (code: ICTS/FTSky2025/10).

This paper makes use of services or code that have been provided by AAO Data Central (datacentral.org.au).
This research has made use of the VizieR catalogue access tool, CDS, Strasbourg, France \citep{Ochsenbein1996}. The original description of the VizieR service was published in \citet{Ochsenbein2000}.

Based on observations made with ESO Telescopes at the La Silla Paranal
Observatory under programme ID 108.21ZF.

Based in parts on observations obtained at the Southern Astrophysical Research (SOAR) telescope, which is a joint project of the Minist\'{e}rio da Ci\^{e}ncia, Tecnologia e Inova\c{c}\~{o}es (MCTI/LNA) do Brasil, the US National Science Foundation’s NOIRLab, the University of North Carolina at Chapel Hill (UNC), and Michigan State University (MSU).

Apart from already meantioned tools, this work made use of \tool{astropy} \citep{Astropy2013}, \tool{dm\_phase} \citep{Seymour2019}, \tool{matplotlib} \citep{Hunter2007}, \tool{psrqpy} \citep{Pitkin2018}, and \tool{seaborn} \citep{Waskom2021}.

          
\section*{Data Availability}

The Data will be made available with the accepted article.



\bibliographystyle{mnras}
\bibliography{frbs} 




\appendix





\bsp
\label{lastpage}
\end{document}